\documentclass[preprint,5p,times,authoryear]{elsarticle}

\usepackage{color}
\usepackage[ruled,vlined]{algorithm2e}
\usepackage{multirow} 
\usepackage{booktabs}
\usepackage{amsmath}
\usepackage{amssymb}
\usepackage[hyphens]{url}  
\usepackage[british]{babel} 

\definecolor{Red}{RGB}{255,37,21}
\definecolor{Green}{RGB}{81,153,74}

\providecommand{\up}[1]{\mathrm{#1}}

\newcommand{\nv}{NVIDIA }
\newcommand{\intel}{Intel }
\begin{document}

\graphicspath{ {figs/}}

\begin{frontmatter}
\title{A polyphase filter for many-core architectures}

\author{Karel Ad{\'a}mek\corref{cor1}}
\ead{karel.adamek@fpf.slu.cz}
\author{Jan Novotn{\'y}\corref{cor1}}
\ead{jan.novotny@fpf.slu.cz}
\address{Institute of Physics, Silesian University in Opava, Faculty of Philosophy and Science, Bezru\v{c}ovo n\'{a}m. 13, 746 01 Opava, Czech Republic}
\author{Wes Armour\corref{cor2}}
\ead{wes.armour@oerc.ox.ac.uk}
\address{Oxford e-Research Centre, University of Oxford, 7 Keble Road, Oxford OX1 3QG, United Kingdom}

\cortext[cor1]{Corresponding author.}
\cortext[cor2]{Principal corresponding author.}

\begin{abstract}
In this article we discuss our implementation of a polyphase filter for real-time data processing in radio astronomy. The polyphase filter is a standard tool in digital signal processing and as such a well established algorithm. We describe in detail our implementation of the polyphase filter algorithm and its behaviour on three generations of \nv GPU cards (Fermi, Kepler, Maxwell), on the Intel Xeon CPU and Xeon Phi (Knights Corner) platforms. All of our implementations aim to exploit the potential for data reuse that the algorithm offers. Our GPU implementations explore two different methods for achieving this, the first makes use of L1/Texture cache, the second uses shared memory. We discuss the usability of each of our implementations along with their behaviours. We measure performance in execution time, which is a critical factor for real-time systems, we also present results in terms of bandwidth (GB/s), compute (GFLOP/s/s) and type conversions (GTc/s). We include a presentation of our results in terms of the sample rate which can be processed in real-time by a chosen platform, which more intuitively describes the expected performance in a signal processing setting. Our findings show that, for the GPUs considered, the performance of our polyphase filter when using lower precision input data is limited by type conversions rather than device bandwidth. We compare these results to an implementation on the Xeon Phi. We show that our Xeon Phi implementation has a performance that is $1.5\times$ to $1.92\times$ greater than our CPU implementation, however is not insufficient to compete with the performance of GPUs. We conclude with a comparison of our best performing code to two other implementations of the polyphase filter, showing that our implementation is faster in nearly all cases. This work forms part of the Astro-Accelerate project, a many-core accelerated real-time data processing library for digital signal processing of time-domain radio astronomy data. 
\end{abstract}

\end{frontmatter}


\section{Introduction}
\label{sec:Introduction}

	The technique of time-domain filtering is a rich and far reaching area in the field of signal processing. One of the cornerstones of time-domain data processing is the use of linear filters. Linear filtering of time-domain signals is a technique employed in many different scientific and industrial settings, from everyday tasks such as audio and video processing to the filtering of radio signals in the field of radio astronomy, it is this latter use of such filters that motivates our work.

	This article focuses on the implementation of a polyphase filter on many-core technologies\footnote{https://github.com/wesarmour/astro-accelerate/tree/master/lib/AstroAccelerate/PPF}. We use the problem posed by real-time signal processing of time-domain radio astronomy data as our application domain, however this work is in no way limited to this field alone. Typical signal processing pipelines in radio astronomy, \cite{scloccoreal} \& \cite{chennamangalam2015artemis} use several processing steps to extract a meaningful signal from input data. This is significant when employing many-core accelerators to achieve real-time processing because (in many cases) it allows data to reside on the accelerator card, circumventing the need for multiple host to device data transfers via the relatively slow PCIe bus. As such our codes can be used in two different ways. The first is as a stand-alone implementation of the polyphase filter, the second is a module that can be incorporated into an existing signal processing pipeline.  Whilst our focus has been to produce implementations that run in real-time, the codes can also be used to process archived data.
	
	Processing time-domain radio astronomy data in real-time enables events, such as Fast Radio Bursts, to be detected and observed as they occur. This allows scientists to create data rich observations by carrying out follow-up measurements whilst an event is occurring. This is a vital part of our effort to understand very rare events \cite{karastergiou2015limits}. However processing such vast data streams produced by modern radio telescopes can be an extremely demanding computational task. Data undergoes many different processes and transformations, such as de-dispersion \cite{armour2011gpu} \& \cite{clarke2014multi}, before a signal from a distant celestial object can be discerned. This is why it is vitally important to ensure that all processes in a data processing pipeline operate as quickly and efficiently as possible, hence the use of many core acceleration \cite{serylak2012observations}, \cite{magro2013multibeam} \& \cite{scloccoreal}.
	
  	In this article we present various implementations of the polyphase filter (PPF). We discuss their application and limitations. We compare the performance of these on several different hardware architectures, three generations of \nv GPUs (scientific and gaming), on CPUs and also the Intel Xeon Phi. We compare our results to two previous works on many-core platforms, the polyphase filter created for the VEGAS spectrometer \cite{PAS:9452994}, and the polyphase filter for the LOFAR radio telescope \cite{vanderVeldt:2012:PFG:2286976.2286986}.
	
	Our work is structured as such: In Section 2 we discuss the polyphase filter and its features, in Section 3 we describe how we have implemented the polyphase filter on our chosen platforms, in Section 4 we discuss the behaviour of GPU implementations in detail and present our Xeon Phi implementation. Section 5 deals with performance comparison with other published work and sample rates per second are presented, we also briefly summarise our experience with different GPU generations. Lastly Section 6 summarises our work.

	
\section{Polyphase filter bank}
\label{sec:Polyphasefilterbank}
\begin{figure}[ht]
		\centering 
			\includegraphics[width=\linewidth]{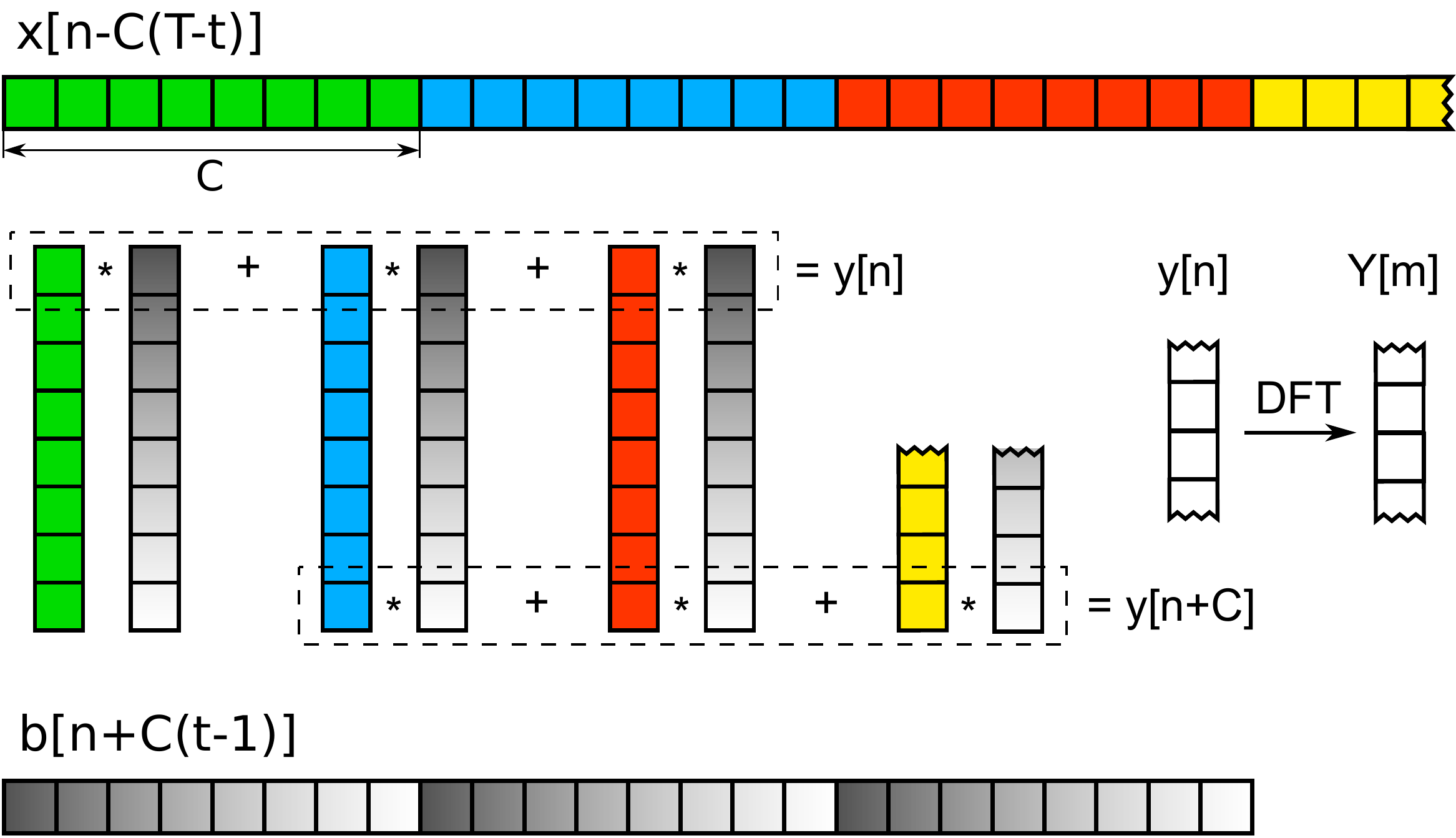}
		\caption{Depicts the structure of an input data stream, the FIR filter operation and DFT. The input data stream (top), is divided into spectra each containing $C$ samples (spectra are differentiated by colour). Each sample (a single square) $x[n]\in\mathbb{C}$ belongs to a specific channel. In this example the spectra have 8 channels. The response function $b$ of the FIR filter must be of size $CT$ (bottom shaded by a gradient). The operation of the FIR filter is shown (middle). The FIR filter takes $T$ number of raw spectra (coloured groups of squares), in this case T=3, and produces one filtered spectra $y[n]$. The next filtered spectra $y[n+C]$ reuses $T-1$ raw spectra from the previous filtered spectra. The DFT acts on filtered spectra $y[n]$ to produce frequency spectra $Y[n]$ (middle, right).}
		\label{fig:DataStructure}
	\end{figure}

	Before we describe the polyphase filter (PPF) algorithm, we shall address the structure of data on which we apply the polyphase filter. We assume input data to be a stream of complex samples $x[n]\in\mathbb{C}$ in the time domain, these are divided into $S$ groups, each containing $C$ samples, we call these groups \textit{spectra}. We refer to the position of samples within a spectra as \textit{channels}. The structure of the input data together with FIR filter is depicted in Figure~\ref{fig:DataStructure}.
	
	The polyphase filter consists of two steps. The first step is to apply a linear filter, which combines $T$ previous time domain spectra, we refer to them as \textit{raw} spectra, into one \textit{filtered} spectra. In the case of the polyphase filter the linear filter combines samples within a single channel of raw spectra and it is described by a finite impulse response (FIR) filter. The second step is to apply a discrete Fourier transformation (DFT) applied on a single filtered spectra which produces a single \textit{frequency} spectra. These two steps are outlined as follows:
	\begin{eqnarray}\label{equ:schemepol}
	\textnormal{raw spectra} \, \xrightarrow{\mathrm{FIR}} \, \textnormal{filtered spectra}\,,\nonumber\\
	\textnormal{filtered spectra} \, \xrightarrow{\mathrm{DFT}} \, \textnormal{frequency spectra}\,. \nonumber
	\end{eqnarray}
	
	The FIR filter is mathematically given \citep{LYONS:UDSP} by 
	\begin{eqnarray}
	y[n]=\sum_{t=1}^{T}x[n-C(T-t)]b[CT-C(t-1)]\,,
	\label{equ:FIR}
	\end{eqnarray}
	where $0 \le n < C$, square brackets $[\,]$ indicate that a physical quantity is discrete (sampled), $x[n]$ represents samples from the input data belonging to the raw spectra and the quantity $y[n]$ represents samples in the filtered spectra. Quantities $y[n]$ and $x[n]$ are assumed to be complex. The FIR filter is a convolution of samples $x[n]$ within a single channel with coefficients of a response function $b$. The number of past samples which the FIR filter operates on is called \textit{taps}, denoted by $T$. The choice of a response function $b$ depends on the desired features of the polyphase filter. The data access pattern for the FIR filter is depicted in Figure~\ref{fig:DataStructure}.
	
\begin{figure}[ht]
		\centering 
			\includegraphics[width=\linewidth]{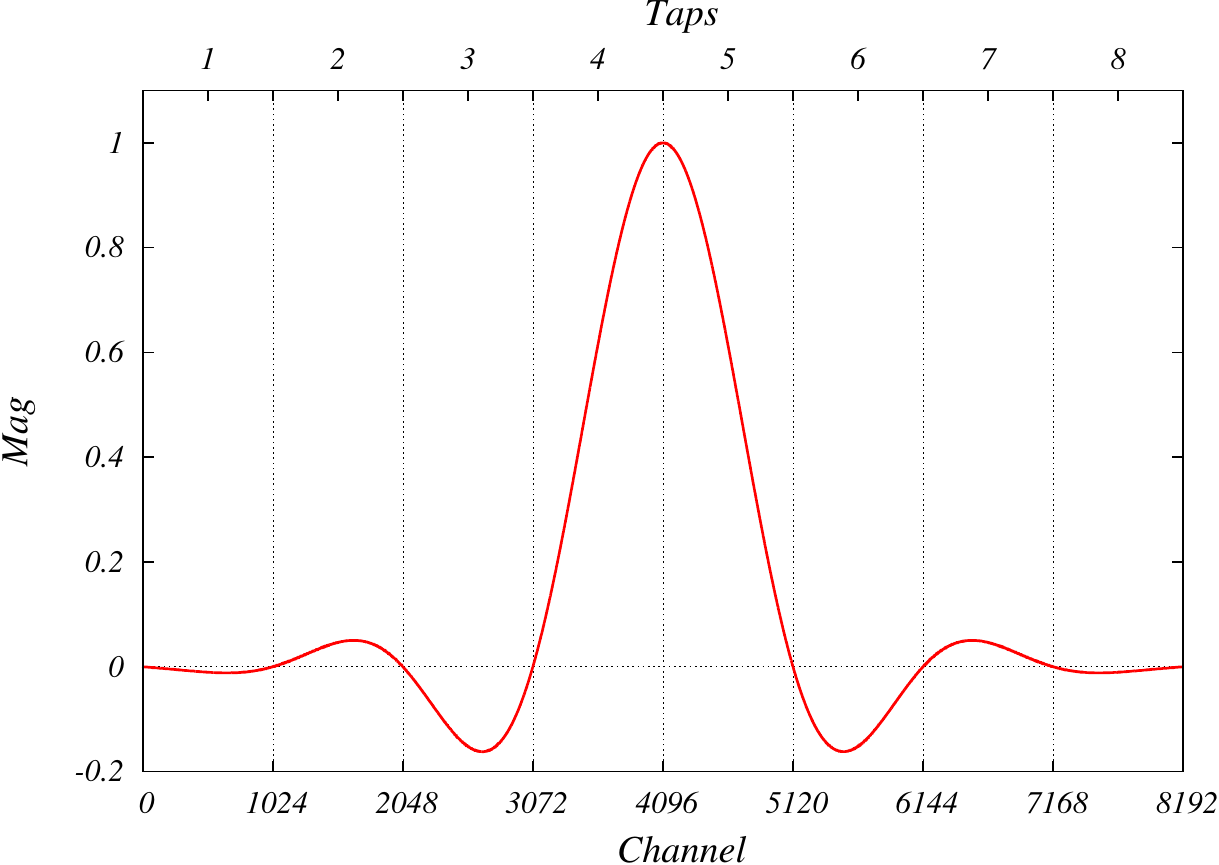}
		\caption{\label{fig:ppf-coeff} Values of the coefficients of the response function $b[n]$ used in examples with 1024 channels per spectra. Coefficients are generated by a $\mathrm{sinc}(x)$ function and multiplied by a Hanning window. On the $x$-axis we have the position within coefficients $0 \le n < CT$, where $CT$ is the total number of coefficients. The division of the coefficients into taps (in this case $T=8$) is depicted by vertical dashed lines.}
	\end{figure}

	We have used a $\mathrm{sinc}(x)$ function to generate the coefficients used in this article, however these are easily replaceable in our code. The $\mathrm{sinc}(x)$ function in the time-domain transforms into a pair of a rectangular windows in the frequency domain. To obtain more accurate results we have multiplied the $\mathrm{sinc}(x)$ function by a Hanning window \cite{LYONS:UDSP}. The resulting coefficients can be seen in Figure~\ref{fig:ppf-coeff}.
	
	The discrete Fourier transformation (DFT), which forms the second step of the polyphase filter, is given \cite{LYONS:UDSP} by 
	\begin{equation}
	Y[m]=\sum_{n=0}^{C-1}y[n]\exp{\frac{-i2\pi nm}{C}},
	\label{equ:FFT}
	\end{equation}
	where $Y[m]$ represents data in frequency domain and $C$ represents the number of channels.
	
\begin{figure}[ht]
		\centering 
			\includegraphics[width=\linewidth]{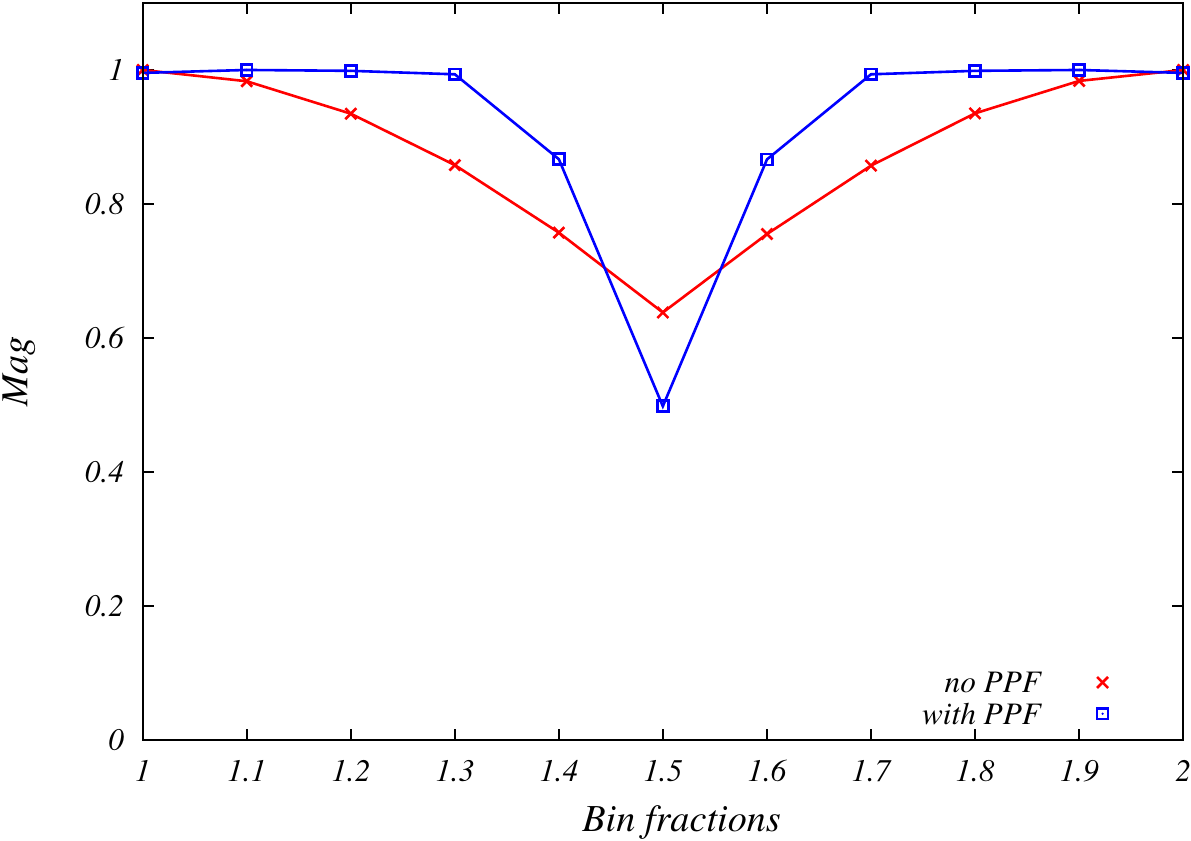}
		\caption{\label{fig:ppf-DFTs} An example of DFT scalloping loss. Scalloping loss is reduced when a PPF is employed.}
	\end{figure}

\begin{figure}[ht]
		\centering 
			\includegraphics[width=\linewidth]{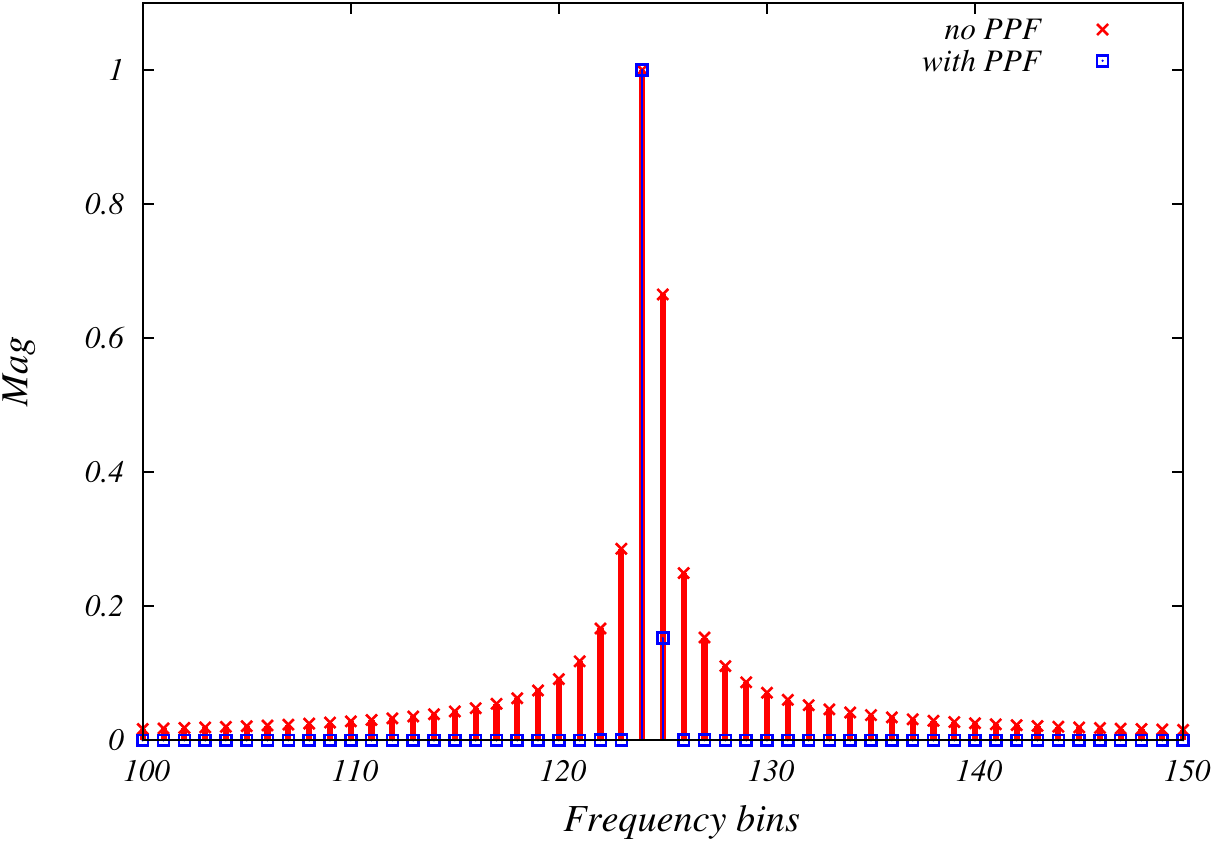}
		\caption{\label{fig:ppf-DFTl} An example of DFT leakage. Leakage is greatly reduced by applying a PPF.}
	\end{figure}

	The polyphase filter reduces errors introduced by a discrete Fourier transformation, these are DFT leakage and DFT scalloping loss, depicted in Figures~\ref{fig:ppf-DFTs} and \ref{fig:ppf-DFTl}. The polyphase filter can also serve for sample rate conversions and as a bandpass filter. Figures~\ref{fig:ppf-DFTs} and \ref{fig:ppf-DFTl} give examples of DFT scalloping loss and DFT leakage respectively. Scalloping loss manifests itself as a magnitude loss centred on frequencies between centres of the DFT bins. By sweeping the frequency-domain with a single tone going from one bin centre (let's say the 1st bin) to a neighbouring bin centre (the 2nd bin) and plotting the maximum magnitude we arrive at Figure~\ref{fig:ppf-DFTs}. It can be observed that DFT scalloping loss has been reduced by the application of a PPF. 
	
	Signal leakage causes a smearing of a signal's frequency into all other frequency bins, this occurs when the frequency is not equal to the central frequency of one of the frequency bins. To produce an example of DFT leakage it is sufficient to generate a single tone with the above mentioned frequency. The DFT leakage and how it is reduced after the application of a PPF can be seen in Figure~\ref{fig:ppf-DFTl}.
	
\section{Implementation}
\label{sec:implementation}
	
	Our primary goal is to produce a polyphase filter which executes as fast as possible. This is because in a real-time signal processing pipeline faster executing code results in more processing time available to other operations in the pipeline. This allows for more complex signal processing pipelines to be constructed or for data to be searched in finer grained detail. We quantify the real-time performance by the number of samples a platform can process per second, the sample rate, $f_{s}$. We also assess each platforms scalability with respect to two parameters, channels and taps, which we hope reflects the processing needs of data streams produced by most modern radio telescopes.
	
\begin{table*}[t]
\centering
\caption{The platforms used in our study. The number of computational units for GPUs is given by number of GPU cores. For CPU or Xeon Phi we assume one lane in the VPU (see sections \ref{lb:cpu} and \ref{lb:phi}) as a computational unit. Hence the number of computational units will depend on number of threads per CPU/Xeon Phi core and is calculated as \textit{(number of physical cores)} $\times$ \textit{(VPU lane width)} $\times$ \textit{(number of threads per core)}. Our choice of the number of threads per CPU/Xeon Phi core is 1 and 2 respectivly, i.e. we do not consider hyperthreading. For our CPU hardware $16\times8=128$ and for the Xeon Phi $61\times16\times2=1952$. Note: For NVIDIA GPUs we use the NVIDIA profiling tool, nvvp, to deterine some of these values.}
\begin{tabular}{l r r r r r r r }
\toprule
\multirow{2}{*}{Platform} & 
Frequency &
Comp. &
Performance &
\multicolumn{2}{c}{Memory} &
Async copy &
PCI-E\\
\cmidrule(r){5-6}
&
(GHz)& 
Units (\#)&
(TFlops) &
\multicolumn{1}{c}{Size (GB)} &  
\multicolumn{1}{c}{Bandwidth (GB/s)} &  
(\#) &
(ver.)\\
\midrule
GTX 580 & 1.63 & 512 & 1.67 & 3 & 196 & 1 & 2.0\\
GTX 780 Ti & 1.07 & 2880 & 6.18 & 3 & 336 & 1 & 3.0\\
GTX 980 & 1.37 & 2048 & 5.60 & 4 & 224 & 2 & 3.0 \\
Tesla M2090 & 1.30 & 512 & 1.33 & 5 & 177 & 2 & 2.0\\
Tesla K40 & 0.75 & 2880 & 4.29 & 12 & 288 & 2 & 3.0 \\
Tesla K80 & 0.82 & 2496 & 4.11 & 12 & 240 & 2 & 3.0\\
Titan X & 1.09 & 3072 & 6.69 & 12 & 336 & 2 & 3.0\\
\midrule
Xeon E5-2650 2x & 2.00 & 128 & 0.51 & x & 102 & x & x\\
Xeon Phi 7110P & 1.10 & 1952 & 2.15 & 8 & 352 & x &3.0\\
\bottomrule
\end{tabular}
\label{tab:hardware}
\end{table*}

	In this work we study three different types of computational hardware commonly used in high performance computing (HPC). These platforms are Intel CPUs, Intel Xeon Phi, and \nv GPUs. Table~\ref{tab:hardware} lists the hardware under consideration. We have also produced optimised codes for three generations of GPU\footnote{\nv~ Fermi\cite{book:fermi-whitepaper}, Kepler\cite{book:kepler-whitepaper} and Maxwell\cite{book:maxwell-whitepaper}}. For each generation we have produced a cache/texture based code (where applicable) and a shared memory based code. We have produced texture cache based codes for the Kepler and Maxwell generations because the read-only texture cache is easily accessible using intrinsic instructions. The texture cache is a fast read-only memory introduced in the Kepler generation. In the Maxwell generation the L1 and texture cache have been merged into a Unified cache. Lastly the shared memory is a user managed cache which allows the user to control what data are resident in the memory. Shared memory has a larger bandwidth than L1/Unified cache, which is important for bandwidth bound applications.
	
	An important characteristic of a computational platform for real-time data processing is the ability to hide computations behind data transfers. By this we mean that we aim to (wherever possible) simultaneously transfer the data needed for the ${i+1}$ polyphase computation whilst we compute the polyphase on data $i$ and transfer the already computed $i-1$ data. That way we can most effectively use the host to device bandwidth available to us. This is considered in Section \ref{sec:Results} and depicted in Figure~\ref{fig:streams}. 

	For GPU and Xeon Phi hardware, data is transported via a PCIe bus, which can (in its 3.0 incarnation) perform bi-directional transfers up to $16$~GB/s. The PCIe bus bandwidth becomes the limiting factor in performance if a polyphase filter is used as a stand-alone code. However this bottleneck must be considered in the context of the expected use of a polyphase filter. The main aim of this work is to produce a real-time polyphase filter to be used in pipelines with more steps than the polyphase filter alone. From this perspective PCIe bandwidth is likely not to be the limiting factor due to the time taken to execute the end-to-end pipeline. 
	
	\subsection{Input, output}
	\label{subsec:InputOutput}
	Input data can have different levels of precision, manifested in the number of bits of individual samples. To reflect this we have produced three versions of all of our GPU implementations, these are 32-bit (stored as \texttt{float}), 16-bit (\texttt{ushort}) and 8-bit (\texttt{uchar}). Often it is beneficial to store lower precision data in tuples of 2 or 4, i.e. \texttt{ushort2} for 16-bit data or \texttt{uchar4} for 8-bit data. Packing elements allows us to transfer more elements per WORD\footnote{The number of bits handled by the processor as a unit. the WORD size may vary on different platforms.} thus using available bandwidth more effectively. 
	
	Data can be aligned in two different ways, the first is contiguous in spectra, the other more natural form is when data is contiguous in channels. All of our implementations assume data alignment to be contiguous in channels. This means that a receiver samples the whole bandwidth of the instrument and forms a single spectra per time sample, giving all channels of the first spectra then all channels of the second spectra and so on.
	
	Our implementation of the FIR filter (\ref{equ:FIR}) assumes that coefficients are unique for each channel and each tap used in the calculation of one filtered spectra. Coefficients are always stored as 32-bit floating point numbers even when we are working with lower precision data (8-bit or 16-bit data samples). The reason for doing this is to minimise the error introduced by the coefficients into the resulting spectra. The filtered spectra are always stored in 32-bit precision as are the resulting frequency spectra.
	
	\subsection{General considerations for implementation}
	\label{subsec:Considerations}
	The polyphase filter consists of two steps as described in the previous section. The first is the FIR filter applied to each channel, the second is a DFT applied to each whole spectrum. In this work we assume that hardware specific fast Fourier transformation (FFT) libraries are well optimised and so we employ the library relevant to our target hardware to perform the DFT step. We have found the most optimal way to perform the FFTs is to used batched execution.
To perform the FFT we use CUDA cuFFT on \nv GPU's and MKL on \intel CPUs and Xeon Phi. Since we use FFT libraries throughout the rest of this work we continue our discussion focusing only on the FIR filter step of the polyphase filter algorithm, since this is the part we can influence the most. 	
	\begin{figure}[ht]
		\centering
			\includegraphics[width=\linewidth]{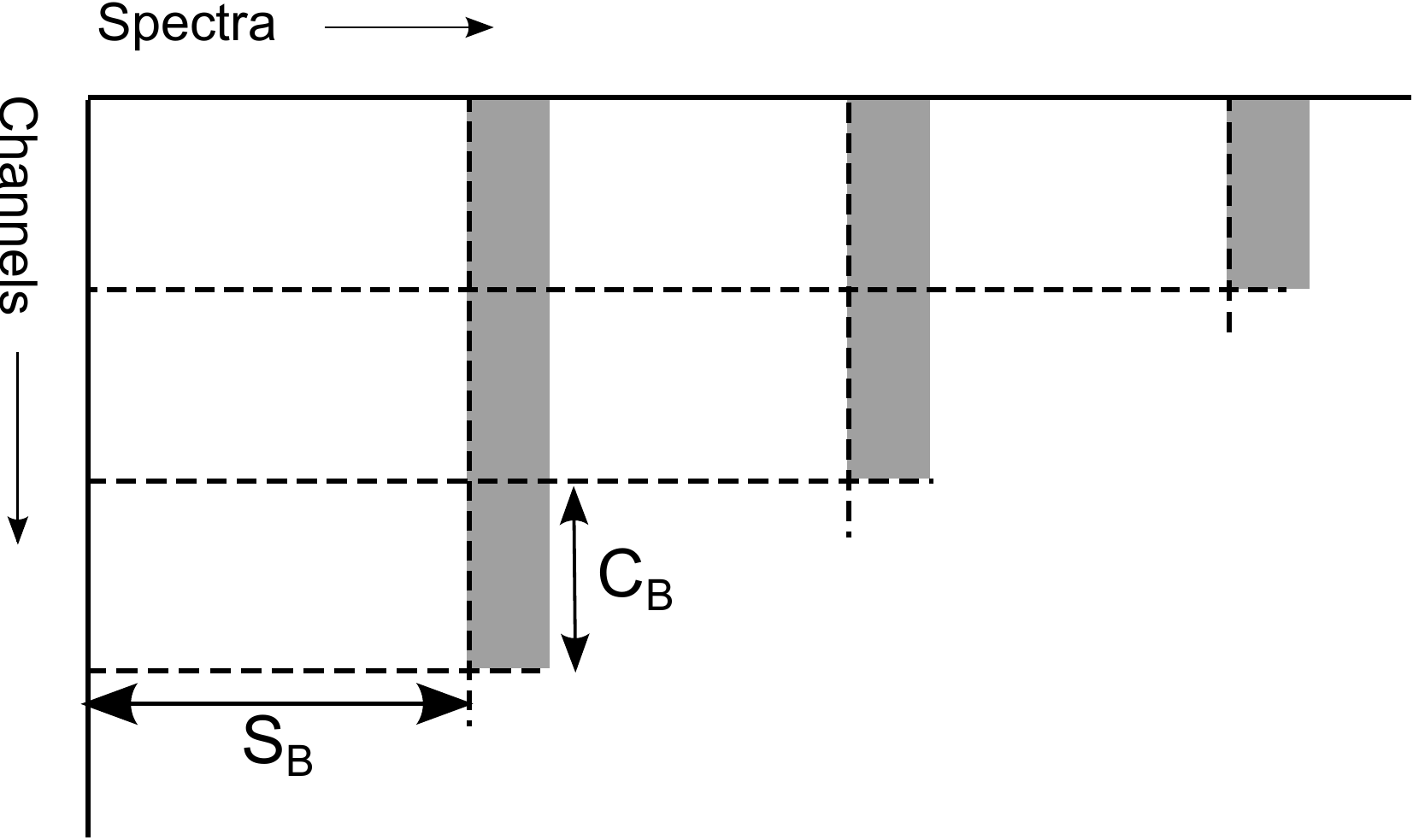}
		\caption{Blocking of input data. This illustration shows how input data are divided into blocks of smaller parts which can then be fitted into faster but smaller memory. Dimensions of each block are given by the number of channels per block $C_\up{B}$ and the number of filtered spectra $S_\up{B}$ calculated from the block. The grey area shows data that is shared by neighbouring blocks.}
		\label{fig:blocking}
	\end{figure}
	
	From a parallelization perspective the FIR filter step poses no resistance. The FIR filter operates on independent channels of multiple spectra, the coefficients used for a FIR filter are the same for each filtered spectra and although raw spectra are reused multiple times, the filtered spectra can be calculated independently of one another. Algorithm~\ref{alg:simplePFB} provides an outline of a serial implementation of the FIR filter step in the polyphase filter.

	This embarrassingly parallel nature of the FIR filter allows us to separate the input data into blocks in both channels and spectra, i.e. we have used a 2D grid. The introduction of blocking in the spectra direction allows instruction parallelism, i.e. we increase thread work load and thus increase the number of memory requests in flight. Increasing in flight memory requests is a well known technique used to increase the fraction of peak memory bandwidth that an algorithm achieves. This is essential in bandwidth limited algorithms like the FIR filter. Our blocking scheme is shown in Figure~\ref{fig:blocking}. The size of the block is given by $C_\up{B}$ number of channels per block and $S_\up{B}$ number of filtered spectra calculated per block, thus the size of a single block is $(S_\up{B}+T-1)\cdot C_\up{B}$ samples. The variable size of these blocks gives us the opportunity to set their size in such a way that they will fit into the fastest memory on the hardware under investigation\footnote{Such as L1 Cache for CPUs or Xeon Phi and Shared memory on GPUs}. The grey areas are data that must be accessed by both blocks that share a common border and has size $(T-1)\cdot C_\up{B}$ samples. Data reuse in the case of the FIR filter is critical since $T-1$ raw spectra used in the calculation of the last filtered spectra are reused for the calculation of the current spectra. Lastly we define a \textit{column} to be a part of the raw spectra contained within the block. In a block of size $C_\up{B}$ one column has $C_\up{B}$ samples. If $C_\up{B}=C$ then one column is equal to one raw spectra.
	
	\begin{algorithm}
 \SetAlgoLined
	\SetKwData{nChannels}{C}\SetKwData{nTaps}{T}\SetKwData{nSpectra}{S}
	\SetKwData{Coeff}{coeff}
	\SetKwData{Data}{input\_data} \SetKwData{Spectra}{spectra}
	
	\SetKwFunction{MLoad}{\_mm\_loadu\_ps}\SetKwFunction{MSetZ}{\_mm\_setzero\_ps}
	\SetKwFunction{MMul}{\_mm\_mul\_ps}	\SetKwFunction{MAdd}{\_mm\_add\_ps}
	\SetKwFunction{MSto}{\_mm\_store\_ps}
	\SetKwInOut{Input}{input}\SetKwInOut{Output}{output}

	\For{$s=0$ \KwTo \nSpectra}{
		\For{$c=0$ \KwTo \nChannels}{
			\For{$t=0$ \KwTo \nTaps}{
				\Spectra[C$\cdot$s+c] = \Coeff[C$\cdot$t+c] * \Data[C$\cdot$(s+t)+c] + \Spectra[C$\cdot$s+c]\;
			}
		}
	}
 
 \caption{Pseudo-code for the serial implementation of a PPF filter without the DFT step.}
\label{alg:simplePFB}
\end{algorithm}

	\label{lb:alg}
	The FIR filter is affected by two parameters, the number of channels $C$ and number of taps $T$. Assuming a fixed number of filtered spectra\footnote{A value of 10 million spectra is used throughout this paper.} $S$ and a fixed number of taps $T$, increasing the number of channels only increases the amount of data which has to be processed, i.e. we have a greater number of blocks in the channel direction and hence performance should scale linearly. Increasing the number of taps $T$ increases number of operations that must be performed by each FIR filter, but also increases the potential for data reuse. Good scaling in the number of taps depends on how effectively we can reuse data, but more importantly how quickly we can reuse the data. The effectiveness of data reuse is improved by increasing $S_\up{B}$. In order to calculate one filtered spectrum we need to load $T$ raw spectra, for the subsequent filtered spectrum we can reuse $T-1$ raw spectra from the previews calculation. Thus we only need to load one new raw spectrum, so saving $T-2$ loads. This is achieved by increasing $S_\up{B}$.

\subsection{GPU Implementations}
	Our GPU implementations form the bulk of our work. We have chosen to study both gaming and scientific cards of the last three generations of \nv GPU architectures. We have used cards from the Fermi (GTX\,580, Tesla m2090 ), Kepler (GTX\,780\,Ti, Tesla K40m, K80) and Maxwell (GTX\,980, Titan\,X) generations. Gaming cards are easily available and cheaper than scientific cards, however they are not designed for sustained computations in a HPC environment.
	Each generation of GPU is different and the implementation of the polyphase algorithm has to be tuned to reflect these differences. As such we have produced CUDA kernels for each version of GPU generation we have investigated. All codes are written using the CUDA programming language and for compilation and profiling we use the CUDA toolkit\footnote{version 6.5 (Fermi, Kepler), we chose to use version 7.0 for the Maxwell architecture because version 6.5 seemed to have stability issues on this platform.}.
	Since the coefficients of a FIR filter are unique for each channel and tap we do not make use of the constant memory present on GPUs. This is because the constant memory uses broadcast, hence when dealing with unique coefficients memory accesses will serialise.

	\subsubsection{Cache Kernel}
		The implementation of our cached based kernel is relatively straightforward. Pseudo-code for our cache kernel is presented in Algorithm~\ref{alg:GPUcache}. In essence one thread block calculates the FIR filter on one block of data shown in Figure~\ref{fig:blocking}. 

		\begin{algorithm}
 \SetAlgoLined
	\SetKwData{nChannels}{$C$}\SetKwData{nTaps}{$T$}\SetKwData{nSpectra}{$S$}
	\SetKwData{CB}{$\up{C_B}$}
	\SetKwData{SB}{$\up{S_B}$} \SetKwData{accumulator}{acc}
	
	\SetKwData{Coeff}{coeff}
	\SetKwData{Data}{input\_data}
	\SetKwData{Spectra}{spectra}
	
	\SetKwFunction{TDfloat}{float}
	\SetKwFunction{TDint}{int}
	
	\SetKwBlock{CudaBlock}{GPU Kernel start}{return}

	\For{$k=0$ \KwTo \nChannels/\CB}{
		\CudaBlock{
			\emph{thread block x-grid direction in spectra}\;
			\emph{defining temporary accumulator}\;
			\TDfloat \accumulator[\SB]\;
			\emph{thread id}\;
			\TDint Th=threadIdx.x\;
			\TDint Bl=\SB$\cdot$C$\cdot$blockIdx.x\;
			\emph{Calculating FIR filter}\;
			\For{$t=0$ \KwTo \nTaps}{
				\For{$s=0$ \KwTo \SB}{
					\accumulator[s] = \Coeff[C$\cdot$t+k$\cdot$\CB+Th] * \Data[C$\cdot$(t+s)+k$\cdot$\CB+Th+Bl] + \accumulator[s]\;
				}
			}
			
			\emph{store to global memory}\;
			\For{$s=0$ \KwTo \SB}{
				\Spectra[C$\cdot$s+k$\cdot$\CB+Th+Bl]=\accumulator[s]\;
			}
		}
	}
 
 \caption{Pseudo-code for our GPU cache kernel implementation of the PPF. The kernel is launched multiple times depending on configuration.}
\label{alg:GPUcache}
\end{algorithm}

		Each row of blocks (all blocks in the spectra direction) are executed as separate kernel launches. This ensures we have some control over the data resident in cache. We found this approach produces faster executing code than using a $y$ grid dimension. This clearly improves scaling in the channels dimension and adds flexibility in terms of the number of channels that the code is able to process. The correct choice for the value of $C_\up{B}$ depends on the number of taps being processed. A large value of $C_\up{B}$ increases \textit{occupancy}, i.e. the number of active thread groups\footnote{Called a \textit{warp} in the CUDA programming model, currently the \textit{warp} size is 32 threads for all CUDA capable hardware.}. However if combined with a larger number of taps higher values of $C_\up{B}$ increases the number of cache misses, this is because each thread block requires more memory space in cache thus limiting data reuse between thread blocks. A small number of $C_\up{B}$ is beneficial for cache utilisation, but decreases occupancy.

		As discussed in section \ref{lb:alg}, to utilise a higher fraction of peak memory bandwidth we use $S_\up{B}$. In the case of our cache kernel this has the adverse effect of increasing register usage per thread and thus decreases occupancy.
	
		The great benefit of our cache based kernels is that they are far less susceptible to the number of taps used, when compared to our shared memory based kernels. The cache kernel also performs reasonably well for almost any configuration of channels and taps.
	
	\subsubsection{Shared Memory Kernel}
		The advantage of using shared memory on a GPU is twofold. Typically a programmer can achieve higher bandwidth when compared to codes that utilise L1 or texture memory alone, also shared memory acts as a user managed cache, this allows the user to ensure the data they wish to operate on is resident in shared memory and hence data reuse can be maximised. This eliminates our cache kernels dependency on the caching behaviour of L1 or texture memory. Here we describe the code for the 32-bit input data case, however where differences occur in the 16-bit and 8-bit cases we highlight these and outline them.
		
		\begin{figure}[ht]
			\centering 
				\includegraphics[width=\linewidth]{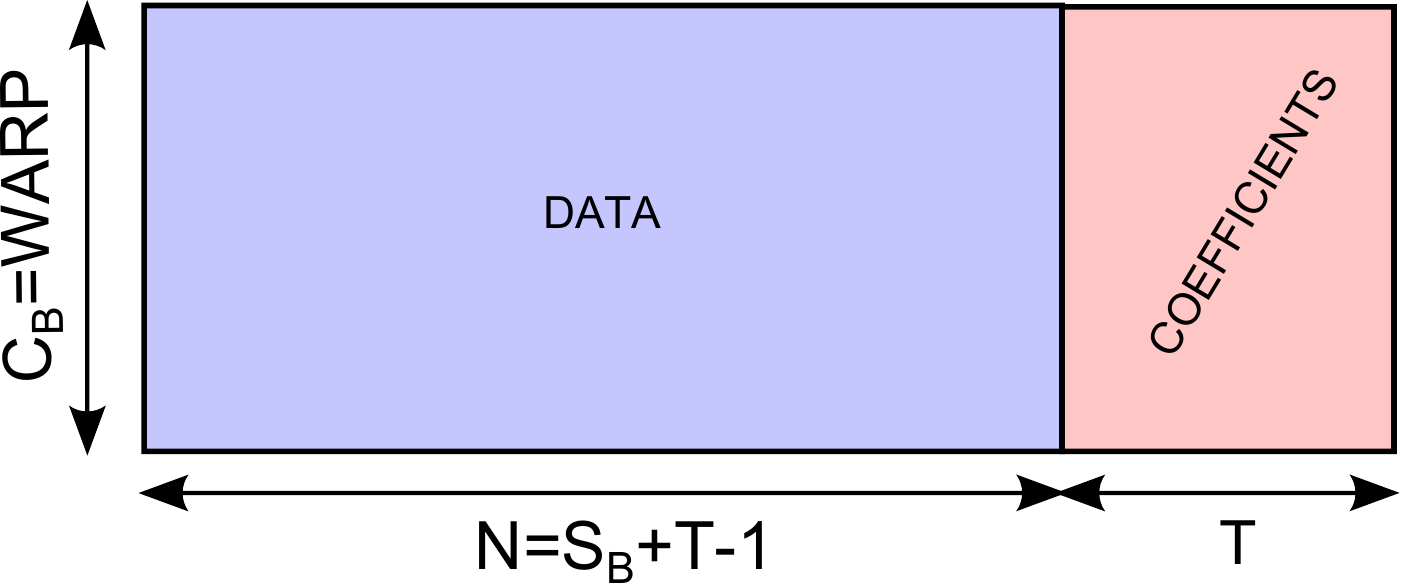}
			\caption{Data structure in shared memory. The first area is dedicated to input data and the second area to coefficients which are reused for each filtered spectra.}
			\label{fig:datastructure}
		\end{figure}
	
		\begin{figure}[ht]
			\centering 
				\includegraphics[width=\linewidth]{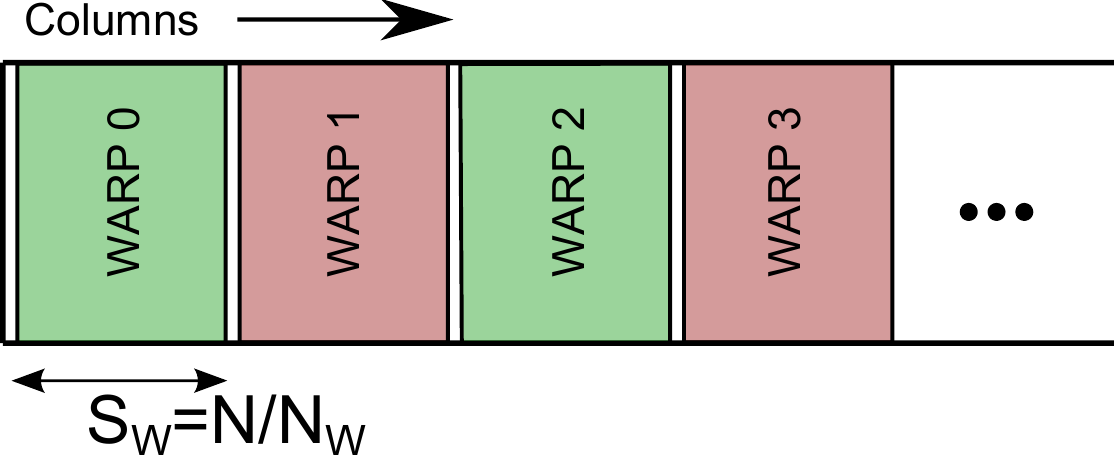}
			\caption{\label{fig:datadivision} Division of data area in shared memory into sub-blocks. Each sub-block is managed by a single \textit{warp}.}
		\end{figure}
		
		The shared memory implementation again works with input data divided into blocks as shown in Figure~\ref{fig:blocking}. We have chosen to split the shared memory available to each thread block into two sections, this is shown in Figure~\ref{fig:datastructure}, one for data and the second for coefficients (which are reused for each filtered spectra). This division along with the restricted amount of shared memory gives an upper limit to the number of taps for which this implementation will work. The maximum number of taps depends on the hardware generation of the GPU and on the bit precision of the input data.

		The data section is then further divided into sub-blocks, shown in Figure~\ref{fig:datadivision}, where each sub block is managed (loading data, computing FIR) by a single \textit{warp}. The number of \textit{warps} per thread block is given by $$N_\up{W}=\frac{(\textit{number\,of\,threads})}{(\textit{size\,of\,\textit{warp}})}.$$ Table~\ref{tab:sharedmemkernel} gives more technical details on the implementations on different generations and bit precision's. Pseudo-code for our shared memory kernel is showed in Algorithm~\ref{alg:GPUSM}.

		\begin{algorithm}
 \SetAlgoLined
	\SetKwData{nChannels}{nChannels}
	\SetKwData{nTaps}{nTaps}
	\SetKwData{nSpectra}{nSpectra}
	\SetKwData{SW}{$\up{S_W}$}
	\SetKwData{SB}{$\up{S_B}\!\!$}
	\SetKwData{WId}{$\up{I_{W}}\!\!$}
	\SetKwData{ThId}{$\up{I_{Th}}$}
	\SetKwData{lThId}{$\up{I_{L}}\!\!$}
	\SetKwData{xsb}{$\up{x_{sb}}\!\!$}
	\SetKwData{xcol}{$\up{x_{cl}}\!\!$}
	\SetKwData{xGl}{$\up{x_{Gl}}\!\!$}
	\SetKwData{yGl}{$\up{y_{Gl}}\!\!$}
	\SetKwData{NW}{$\up{N_{W}}\!$}
	\SetKwData{dts}{$\cdot$\!\!}
	\SetKwData{accumulator}{acc}
	
	\SetKwData{GlCoeff}{Gl\_coeff}
	\SetKwData{SMCoeff}{SM\_coeff}
	\SetKwData{Spectra}{Gl\_spectra}
	\SetKwData{GlData}{Gl\_data}
	\SetKwData{SMData}{SM\_data}
	
	\SetKwFunction{TDfloat}{float}
	\SetKwFunction{TDint}{int}
	\SetKwFunction{GLmem}{LoadData}
	
	\SetKwBlock{CudaBlock}{GPU Kernel start}{return}
	\SetKwBlock{LoadBlock}{Loading data}{}
	\SetKwBlock{ComputeBlock}{Computation}{}
	
	\CudaBlock{
		\emph{Grid coordinates and block's dimension}\;
		\TDint x=blockIdx.x\;
		\TDint y=blockIdx.y\;
		\emph{position of current block in \GlData or \Spectra}\;
		\TDint \xGl = x\;
		\TDint \yGl = y\dts\SB+\xcol\,\;
		\emph{WARP size}\;
		\TDint W=32\;
		\emph{Id of a thread, thread's WARP and thread's Id within its WARP}\;
		\TDint \ThId = threadIdx.x\;
		\TDint \WId = (int) (\ThId/W)\;
		\TDint \lThId = \ThId-\WId\dts W\;
		\emph{begining of the WARP's sub-block in shared memory}\;
		\TDint \xsb = \WId\dts\SW\;

		\LoadBlock{
			\emph{each WARP load data into it's sub-block}\;
			\For{$s=0$ \KwTo \SW}{
				\emph{position of current column in \SMData}\;
				\TDint \xcol = s + \WId\dts\SW\;
				\SMData[\xcol$\cdot$W+\lThId] = \GlData[\yGl$\cdot$C + \xGl$\cdot$W + \lThId]\;
			}
			\emph{then each WARP load one or more columns of coefficients depending on number of WARPs and taps}\;
			\For{$t=0$ \KwTo T/\NW}{
				\emph{current column in coefficients}\;
				\TDint \xcol = \WId + t\dts(T/\NW)\;
				\SMCoeff[\xcol$\cdot$W+\lThId] = \GlCoeff[\xcol$\cdot$C+x$\cdot$W+\lThId]\;
			}
		}
		
		\ComputeBlock{
			\emph{defining temporary accumulator}\;
			\TDfloat \accumulator\;
			\emph{Calculating FIR filter}\;
			\For{$s=0$ \KwTo \SW}{
				\TDint \xcol = s + \WId\dts\SW\;
				\uIf{\xcol$<S_B$}{
					\For{$t=0$ \KwTo \nTaps}{
						\accumulator = \SMCoeff[t$\cdot$W + \lThId] * \SMData[\xcol\dts W + \lThId + t$\cdot$W] + \accumulator\;
					}
					\emph{store to global memory}\;
					\Spectra[\yGl$\cdot$C + \xGl$\cdot$W + \lThId]=\accumulator\;
				}
			}
		}
		
	}
 
 \caption{Pseudo-code for our GPU shared memory kernel implementation of the FIR filter. Preposition \texttt{Gl\_} means that data are stored in global memory, \texttt{SM\_} represents the shared memory. Our kernel uses a CUDA grid with channels in the x direction and spectra in the y direction.}
\label{alg:GPUSM}
\end{algorithm}

		The shared memory kernel consists of a loading phase and a computing phase. In the loading phase a \textit{warp} loads all required data into it's memory sub-block and loads it's part ($T/N_\up{W}$) of the coefficients required for the calculation of the FIR filter. Data are read from global memory in the direction of channels, which is the direction of data alignment. The number of data columns loaded depends on the size of the shared memory assigned to each thread block. The resulting global memory accesses are aligned and coalesced.

		\begin{table}
			\caption{Specific details of our shared memory kernel for each hardware generation and input data bit precision. Columns description: \textit{Cache-lines read} is the number of cache-lines (128 B) read from global memory and stored into shared memory. This gives the size of the working data set. \textit{Cache-lines written} is the amount of data written into global memory. The discrepancy between cache-lines read and written is due to the bit precision of the input data (bit precision varies) and output (always 32-bit) data. \textit{Max taps} gives maximum number of taps with which a kernel can efficiently compute the FIR filter. \textit{Threads reads} gives the format of the input data.}
			\centering
			\resizebox*{\linewidth}{!}{%
			\begin{tabular}{llrrrr}
			\toprule
			\multirow{2}{*}[-2pt]{Generation} & \multirow{2}{*}[-2pt]{Precision} & \multicolumn{2}{c}{Cache-lines} &  \multirow{2}{*}[-2pt]{Max taps} & Thread \\
			 \cmidrule(r){3-4} & 											& Read & Written & & 			reads \\
			\midrule
			\multirow{3}{*}[-2pt]{Fermi}	& 32-bit & 2 & 2 & 128 & \texttt{float2} \\
											& 16-bit & 1 & 2 & 192 & \texttt{short2} \\
											&  8-bit & 1 & 4 & 128 & \texttt{uchar4} \\
			\multirow{3}{*}[-2pt]{Kepler} 	& 32-bit & 2 & 2 & 96  & \texttt{float2} \\
											& 16-bit & 1 & 2 & 192 & \texttt{short2} \\
											&  8-bit & 1 & 4 & 128 & \texttt{uchar4} \\
			\multirow{3}{*}[-2pt]{Maxwell} 	& 32-bit & 2 & 2 & 128 & \texttt{float2} \\
											& 16-bit & 1 & 2 & 192 & \texttt{short2} \\
											&  8-bit & 1 & 2 & 192 & \texttt{uchar2} \\
			\bottomrule
			\end{tabular}
			}  
			\label{tab:sharedmemkernel}
		\end{table}

		In the calculation phase a \textit{warp} calculates the FIR filter within it's sub-block. This allows us to utilise all available threads which can cooperate within the thread block. If the number of taps are greater than the size of the sub-block $S_\up{W}$ some \textit{warps} will idle because they do not have all of the necessary data to perform the FIR filter computation. This effect is amplified by an increasing number of taps.
		
		The shared memory kernel has two parameters. The first controls the amount of shared memory per thread block and the second is the number of threads per thread block. In general it is best to have as many threads per thread block as possible without sacrificing occupancy. The same argument holds for the amount of shared memory. We have used the practice of setting a dummy parameter, which is the number of active thread blocks per multiprocessor.  The available shared memory and maximum number of threads available per multiprocessor is divided by the number of active thread blocks giving us the shared memory and number of threads per thread block. This method generally produces well performing configurations. However in cases where we wish to produce more memory requests in flight, it is good to favour the number of threads per thread block, thus not utilising all of the shared memory available.
		
		Some kernels have a higher number of maximum taps than others. This is due to the size difference between the input data and coefficients. The input data bit precision changes while the coefficients used are always 32-bit. This also depends on number of loaded data elements. For example our 16-bit kernel stores data in \texttt{ushort2} thus for input data storage we need only 4 bytes and 4 bytes for coefficients. Hence we save 4 bytes of memory when compared to our 32-bit kernel. Another example is our 8-bit kernel working with \texttt{uchar2} input data. We load data as \texttt{uchar2} from global memory but store it in an interleaved manner in shared memory using a \texttt{uchar4} vector. Hence we use 4 bytes of memory to store 2 bytes of data, this however has the effect of preventing bank conflicts for subsequent loads from shared memory, allowing us to access a greater fraction of peak shared memory bandwidth. So in comparison with 16-bit kernel we save nothing. Working with \texttt{uchar4} means loading two times number of data elements, which requires more coefficients to be present to complete the computation. In this case we need 4 more bytes than in the 16-bit case. Combinations like these result in number of maximum allowable taps showed in Table~\ref{tab:sharedmemkernel}.

\subsubsection{Shuffle Kernel}
	One of the new features of the Kepler generation is a shuffle instruction. The shuffle instruction allows one thread of a \textit{warp} to read the value of a variable of another thread within the \textit{warp}. 
		
		We have produced multiple kernels that have utilised the shuffle instruction, but ultimately all shuffle implementations suffered from unsuitable data, which we assume to be contiguous in channels. This is further discussed in subsection \ref{subsec:InputOutput}. To calculate the FIR filter the \textit{warp} would have to contain samples of the same channel, but from different spectra. This would result in unaligned reads with a stride size of the number of channels $C$. Data which is contiguous in spectra is much more suitable for kernels which make use of the shuffle instruction.

\subsection{CPU Implementation}
\label{lb:cpu}
	Our CPU platform is a dual Intel Xeon E5-2650 server. To achieve high performance on modern CPUs it is essential to utilise the vector processing units (VPU) and use all processor cores available. The VPUs allow the CPU to execute single instruction multiple data (SIMD) operations. The number of operations which may be performed simultaneously by each VPU depends on the generation of the CPU. The Intel Xeon E5-2650 have AVX units\footnote{More at https://software.intel.com/en-us/articles/introduction-to-intel-advanced-vector-extensions} which have 256 bit wide vector registers for SIMD operations. The number of elements that can fit into one VPU register depends on size of individual elements and the the VPU vector width, e.g. for single precision numbers the VPU lane width is $256/32 = 8$. For parallelization across cores we use OpenMP. The CPU code is based on the same blocking scheme as was used in our GPU implementations (Fig.~\ref{fig:blocking}). Each block is assigned to a thread which performs the FIR filter. $C_\up{B}$ is chosen as an integer multiple of VPU register size. Once the FIR filter completes the code launches the FFT to calculate the output frequency spectra.

\subsection{Xeon Phi}
\label{lb:phi}
	The Xeon Phi is a computational accelerator produced by Intel. The Phi is derived from CPU cores and thus shares lot of similarities with the CPU. The Xeon Phi has wider SIMD registers (512 bit) and many more cores. The Knights Corner generation of the Xeon Phi co-processor uses the Intel Many-Core Instruction set (IMCI). A full AVX-512 instruction set will be available in the next generation of Xeon Phi (KNL, or Knights Landing). The programming model is designed to be very similar to that of CPUs and hence this makes Phi attractive to those programmers who are used to working on multi-core CPUs and makes porting existing CPU code to Xeon Phi relatively straightforward. General code can run directly in native mode, assuming the input and output are changed to suite Xeon Phi. Porting code with intrinsic instructions is more problematic since by using intrinsic instructions one binds the code to a specific CPU generation, which in turn imposes constrains on data division. This was proven to be valid by \citep{Sch-Ule:2012:EarlyExp} on wide range of codes or by \citep{Cra-Sch:2012:Openmp} when they investigated the performance of Sparse-Matrix-Vector-Multiplication. 
	
	Again we base the Xeon Phi code on the blocking scheme shown in the Figure~\ref{fig:blocking}. Our Xeon Phi implementation is similar, but not the same as our CPU implementation. The difference between these two platforms arises in how we utilise the cache. Typically efficient Xeon Phi code requires there to be at least two physical threads per core (up to four threads if using hyperthreading). This however reduces the cache size available to each thread. Given this we have found that the most beneficial way to utilise Phi is to to assign the same data block from our CPU implementation to a phi core rather than to an individual thread. We bind threads to a core and these threads then cooperate on computing the FIR filter. Our Xeon Phi implementation is for 32-bit input data only. This is because intrinsic instructions for manipulating lower precision data are not supported by the current Knights Corner generation of Xeon Phi.

\subsection{Ease of use}
	All of our implementations have required different levels of knowledge of both hardware and programming models to complete. It is worth noting that our most advanced GPU implementations have taken significantly more time to develop when compared to our CPU implementation, they have also required a greater depth of knowledge when compared to our CPU or Xeon Phi implementations. Our Xeon Phi implementation was the easiest to complete. This is largely because the hardware differences between CPUs and Xeon Phi are smaller when compared to the differences between GPUs and CPUs. Because we had written our CPU code in a parameterised way, used OpenMP and Intel Intrinsics, allowed us to port our CPU code to Xeon Phi in a few hours. We believe that these significant differences in both development time and levels of knowledge required, should be taken into account when viewing our results section. Specifically when considering the overall cost of both capital and development effort to a project.

	
\section{Behaviour}
\label{sec:Implementationbehaviour}
	In this section we describe the behaviour, performance characteristics and limitations for each of our implementations. We have analysed the behaviour of our implementations on the Xeon Phi coprocessor and on each generation of GPU, however for the sake of brevity we describe the behaviour of our GPU implementation only on the newest Maxwell generation of GPU hardware. Differences within GPU cards of the same generation and compute capability are mainly in the number of SMs\footnote{Streaming multiprocessor - a collection of CUDA cores}, memory bandwidth, memory capacity or clock frequency. 

	\subsection{Preliminaries}
\begin{figure}[ht]
			\centering 
			\includegraphics[width=\linewidth]{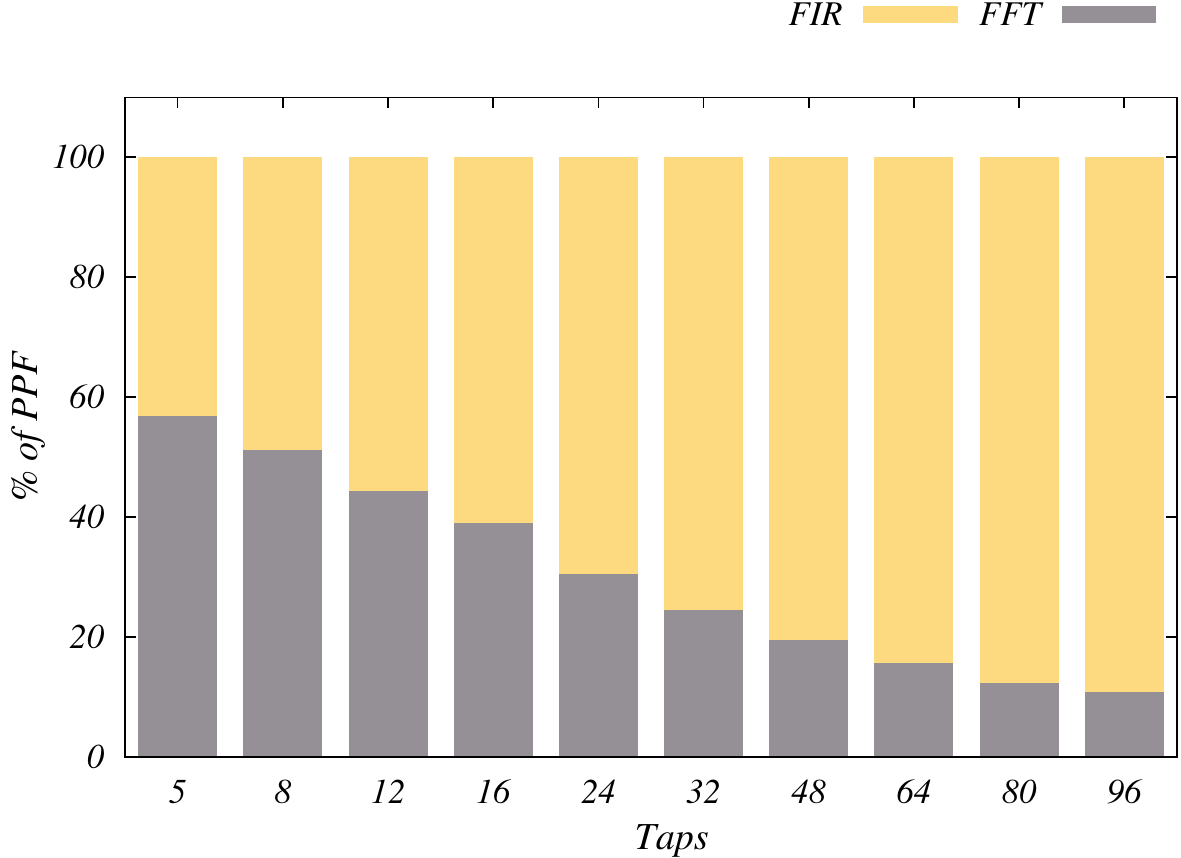}
			\caption{\label{fig:FIRvsFFT} The portion of PPF execution time taken by the FIR and FFT steps respectively, along with the dependency on the number of taps used. We have used 1024 channels in these comparisons. Execution times showed are for 8-bit precision, this is where the FIR filter has best performance, thus the portion of execution time taken by the FIR filter compared to the FFT step is lowest. For higher bit precision's the FIR step becomes even more significant.}
		\end{figure}

		When discussing the behaviour and details of our implementation of the PPF we limit ourselves to discussion of the FIR filter implementation alone (as mentioned in Section \ref{subsec:Considerations}). This is for two reasons, the first is that we employ standard libraries to perform the FFT step, as do other codes that we compare to. The second is that the FIR filter dominates the total execution time of the PPF for a large number of taps. The percentage of execution time taken by the FIR and FFT steps to complete the PPF is depicted in Figure~\ref{fig:FIRvsFFT}, showing that even in the case of a relatively low number of taps (16 taps) the FFT takes only 40\% of total execution time. The final results that we present are for the whole PPF and so include the FFT step, but a discussion here of the FFT implementation would obscure our results, also FFT implementations are sufficiently described elsewhere\footnote{http://docs.nvidia.com/cuda/cufft/index.html, https://software.intel.com/en-us/articles/the-intel-math-kernel-library-and-its-fast-fourier-transform-routines/}. 

		We use 1024 channels for our comparisons (unless otherwise stated), however our implementations are not limited to this specific number. There are some constrains on the number of input spectral channels, they must be divisible by 32 or in some cases 64. In order to have a more precise understanding of which resources are limiting our implementations and how much of various GPU resources different implementations use we compare values reported by the NVIDIA visual profiler (nvvp) with theoretical values. The theoretical shared memory bandwidth is given by
		\begin{equation}
		B_{\mathrm{SM}}=\mathrm{(\textit{warp size})}\times\mathrm{(\#SMs)}\times\mathrm{(Bank\,size)}\times\mathrm{(Core\,freq.)}
		\nonumber
		\end{equation}
		In the cases where a theoretical value cannot be calculated as in the case of L2 and texture cache we have used our own benchmarks. Our benchmark code has been tuned so that \nv profiler reports the highest bandwidth value possible with that code. In the case of the GTX\,980 the profiler shows 85\% utilisation of L2 cache and 95\% utilisation of texture cache. We have also written a benchmark for number of type conversions that can be performed by a GPU. We summarise our findings in Table~\ref{tab:benchmarks}. We use these values in all of our subsequent analysis and plots. In the case of the GTX\,780\,Ti we measured a low performance for type conversions compared to the theoretical value, our measured performance is only 22\% of theoretical peak. The reason for this low performance is unclear.
		
		\begin{table*}
			\centering
			\caption{The theoretical peak or measured bandwidth for each card given by the NVIDIA nvvp profiler. Global memory values are given by a cards specifications. The GTX\,580 theoretical shared memory bandwidth is half of calculated value since each bank can deliver 32 bits per two clock cycles. Bit-conversions are in type conversions per second (GTc/s), we list the percentage of theoretical peak calculated from values in CUDA programming guide section 5.4 in brackets next to these values. The fraction of the theoretical value is very low in the case of the Kepler GTX\,780\,Ti.}
			\begin{tabular}{lrrrrr}
			\toprule
			Hardware & Shared m. [GB/s] & Global m. [GB/s] &  L2 [GB/s] & Texture [GB/s] & Bit-conversions [GTc/s] \\
			\midrule
			GTX 580		& 1669 & 197 & 350  &  NA  & 272 (65\%)\\
			GTX 780 Ti 	& 4116 & 336 & 1300 & 1890 & 441 (22\%)\\
			GTX 980		& 2799 & 224 & 636  & 1412 & 625 (90\%) \\
			\bottomrule
			\end{tabular}
			\label{tab:benchmarks}
		\end{table*}

		To help us identify places where kernel limitations occur, we calculate the performance as \textit{effective FLOPs} per second, i.e. we manually sum the operations a given kernel will perform. Any limiting factors due to GPU resources will appear as a plateau in our plots as we change task parameters. This is because we are either limited by bandwidth - i.e. we cannot get the necessary data to processing cores quickly enough, by compute - i.e. we cannot process the data as quickly as it is delivered or by another resource, for example type conversions. Once a limiting factor is reached, increasing the workload on the compute device will not result in an increase in performance because there are no more available resources to draw upon. These limiting factors can be also caused by a sub-optimal implementation. Our estimation of effective FLOPs do not represent actually executed instructions and do not depend on the actual implementation of the algorithm since they are derived from equation~(\ref{equ:FIR}). This allows us to compare devices independently on specifics of their implementations or compare different implementations on same device. The number of effective FLOPs for a FIR filter given by equation~(\ref{equ:mathFlops}).
		\begin{equation}
		F_\mathrm{FIR}=\frac{2SCT+2SC(T-1)}{t_\up{ex}}=2\frac{SC(2T-1)}{t_\up{ex}},
		\label{equ:mathFlops}
		\end{equation}
		where $F_\mathrm{FIR}$ is the number of FLOPs per second, $S$ is the number of filtered spectra calculated, $C$ is the number of channels, $T$ is the number of taps and finally $t_\up{ex}$ is the execution time of a FIR filter. The factor of two in the fraction takes into account the complex nature of the input data and the factor of two outside the fraction represents one multiplication and one addition which is performed per tap. The calculation of effective FLOPs depends on execution time $t_\up{ex}$ and on volume of data processed. For example a 10 tap FIR filter producing one filtered spectra with 100 channels with execution time $t=1\mathrm{s}$ gives $F_\mathrm{FIR}=3800\,\mathrm{FLOPs}$. The number of instructions that are actually executed could be a half of that number, for example, the hardware might make use of a single FMA instruction which we count as two instructions.
		
		The FIR filter has the potential to be very efficient due to its high data reuse. We quantify this efficiency of a code by evaluating the ratio $\epsilon=B_\mathrm{w}/B_\mathrm{r}$, where $B_\mathrm{w}$ is the amount of data written to memory\footnote{memory assumed is the global memory of the card}, $B_\mathrm{r}$ is the amount of data read from the memory and $\epsilon$ is efficiency of the code. The efficiency in the case of the ideal polyphase filter is\footnote{assuming the number of spectra $S\rightarrow\infty$ for finite number of spectra the efficiency is given by $\epsilon=B_\mathrm{w}/B_\mathrm{r}=S/(S+T-1)$} $\epsilon=B_\mathrm{w}/B_\mathrm{r}=1$. In the ideally in-efficient case, the efficiency is $\epsilon=1/\mathrm{T}$, where $\mathrm{T}$ is the number of taps used. This means that for each filtered spectra, we read all necessary raw spectra from scratch.
		
\begin{figure}[ht]
			\centering 
			\includegraphics[width=\linewidth]{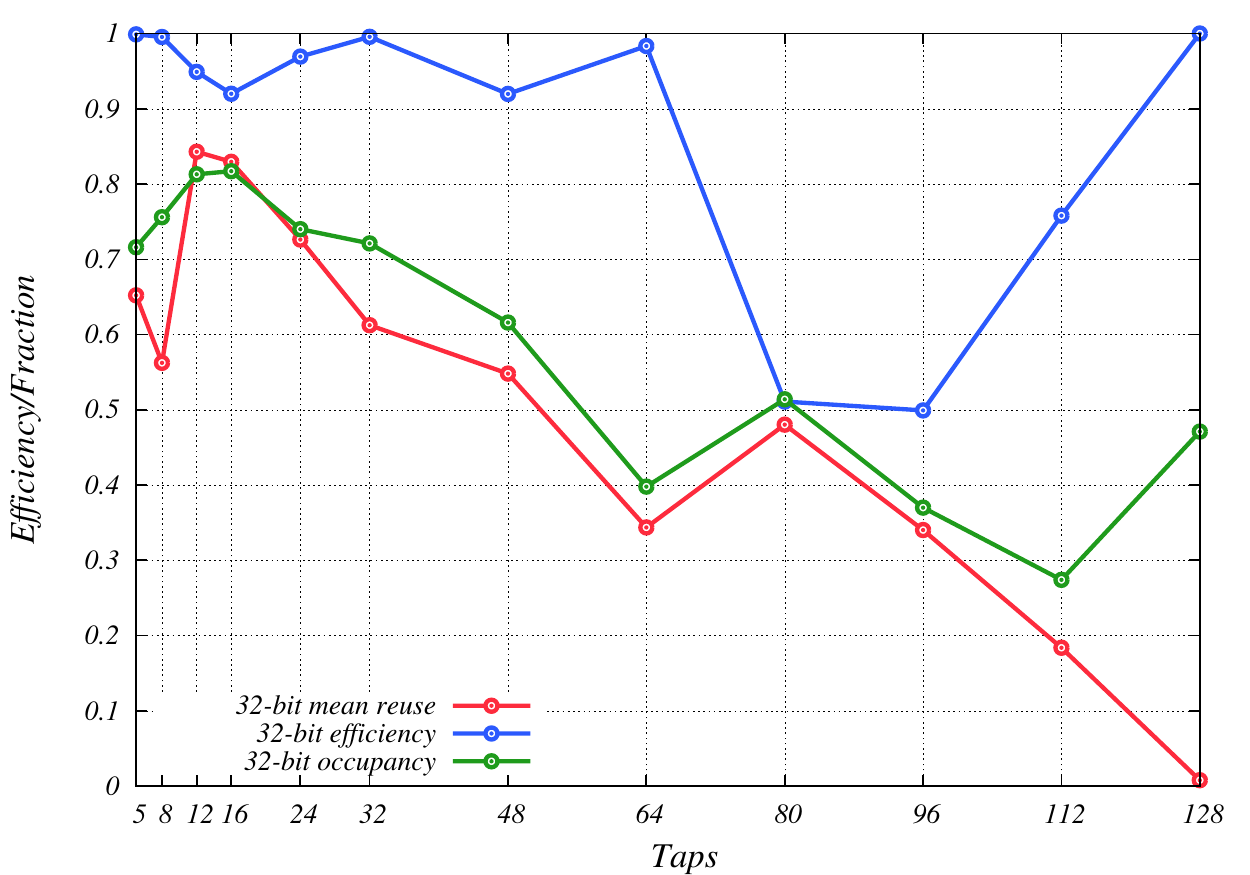}
			\caption{The mean data reuse $R_\up{T}$, efficiency $\epsilon$ and the achieved occupancy as reported by \nv profiler for the best performing configurations. The occupancy follows the mean data reuse closely because we plot values for the best performing configurations and as such an increase in mean data reuse (or occupancy) will lead to a decrease in occupancy (or mean data reuse) and thus a decrease in performance. Exceptions can be observed for both low and high numbers of taps. In both instances there is not enough data reuse in the FIR filter hence mean data reuse is less indicative of performance. Interesting behaviour can be observed at 64 taps, high efficiency is observed indicating that data was not transferred from global memory implying that a decrease in mean data reuse is compensated for by the cache.}
			\label{fig:meanreuse}
		\end{figure}

		The shared memory kernel can be better understood if we introduce a measure of mean data reuse per channel $R$. Since the shared memory space is limited we can only fit $N$ columns of raw spectra data into it, which is given by the relation $N=(\mathrm{Allocated\,memory})/2*4*32=S_{B}+T-1$. The factors 2 and 4 come from the fact that the data is complex (single precision) and the factor 32 because \textit{warp} loads 32 channels\footnote{See section \ref{sec:implementation}}. Data reuse is not the same for each column. We load $N$ raw spectra but we produce only $N-T+1$ filtered spectra, meaning the total data reuse of the first $n=T-1$ columns will increase by one and the data reuse of last $n$ columns will decrease by one. The mean reuse $R$ can be calculated from a series given by the number of data reused in each column
		\begin{equation}
		R=\frac{1+\cdots+(n-1) + n + T + \cdots + T + n + (n-1) + \cdots + 1}{N}.
		\end{equation}
		The leading and trailing series can be easily calculated, since they are arithmetic series given by
		\begin{equation}
		S_n=\frac{n(1+n)}{2}.
		\end{equation}
		The number of fully reused columns is $N-2n$. Putting all this together we arrive at the expression
		\begin{equation}
		R=\frac{(n+1)(N-n)}{N}.
		\end{equation}
		This quantity is however hard to compare with an increasing number of taps, this is why we divide it by number of taps $T$ giving us a value for mean data reuse per column per tap
		\begin{equation}
		R_\mathrm{T}=\frac{R}{T}=\frac{(n+1)(N-n)}{NT}=\frac{N-n}{N},
		\end{equation}
		since $n+1=T$. This number changes with the configuration parameters of the shared memory kernel and together with occupancy has the greatest influence on performance of a given configuration. We have plotted mean data reuse $R_{T}$ and occupancy for the best performing configurations in Figure \ref{fig:meanreuse}. Both quantities depend on the amounts of allocated shared memory per thread block. Mean data reuse will increase with greater shared memory per thread block, but the occupancy will decrease. Hence to achieve the best performance we must find the optimal configuration that allows both mean data reuse and occupancy to be roughly equal and have high values. There is also a fixed amount of shared memory that we must allocate per thread block which is given by number of taps. As this increases it has the effect of decreasing occupancy and mean data reuse at the same time. As the number of taps increases each thread block requires more shared memory to process each FIR filter, this leads to a decrease in occupancy and the possibility for data reuse due to the finite amount of shared memory resources.
 
\begin{figure}[ht]
			\centering 
			\includegraphics[width=\linewidth]{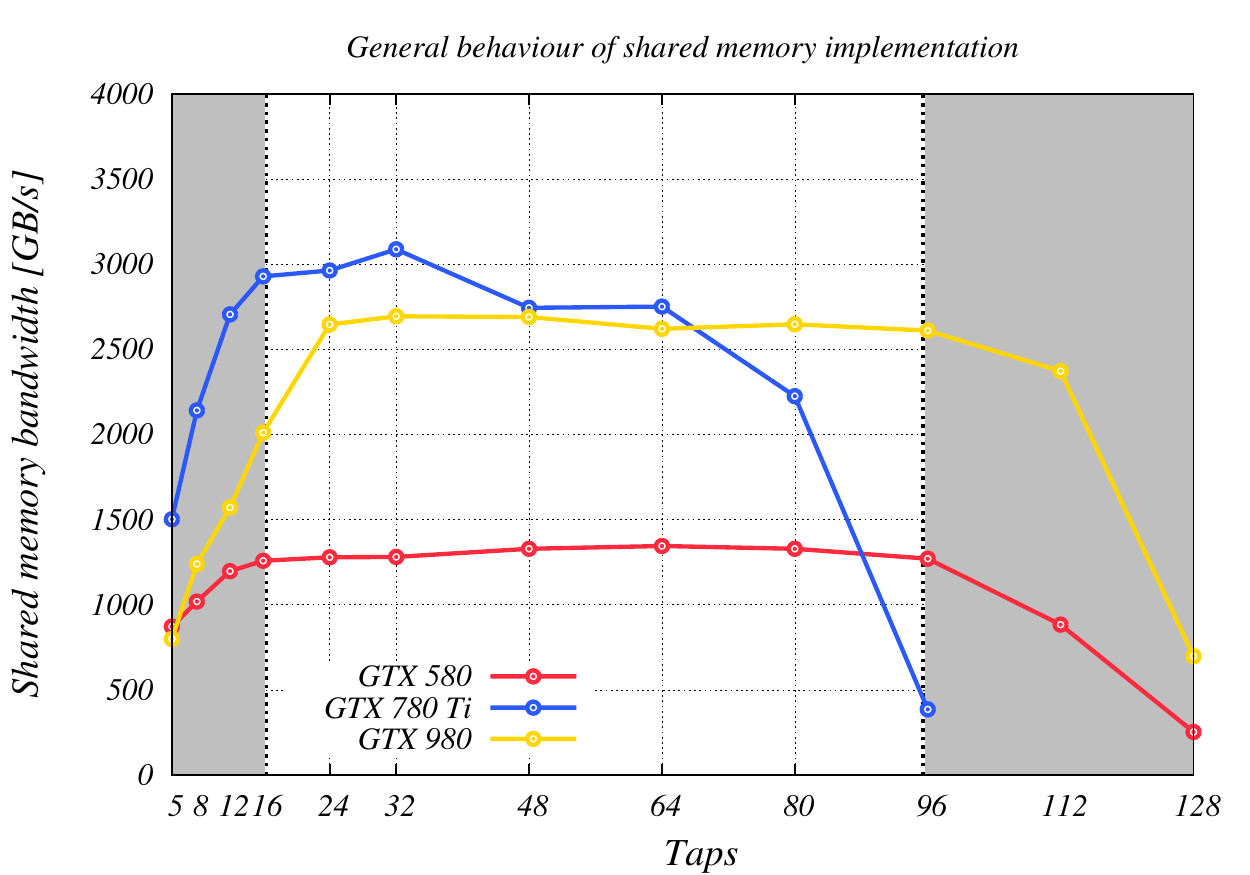}
			\caption{\label{fig:SMGenBehaviour} The general behaviour of our shared memory kernels, with three regions of distinct behaviour change. These are; Global memory bandwidth bound (first grey region), where there is not enough data reuse in the FIR filter and so processing cores cannot be supplied with data quickly enough from global memory. Shared memory bandwidth bound (white plotted region), here enough data reuse occurs to maximise bandwidth to data stored in shared memory. Shared memory capacity bound (third grey region), this is where the shared memory no longer has enough space to enable data reuse and to support enough active thread blocks to hide execution dependencies. Due to implementation differences some implementations might have a shorter middle phase, this can be observed in the data from the GTX\,780\,Ti).}
		\end{figure}

		Lastly we explore the generalised behaviour of our shared memory and cache kernels. The shared memory kernel's performance will, in general, follow the curve shown in Figure~\ref{fig:SMGenBehaviour}. Figure~\ref{fig:SMGenBehaviour} is separated into three regions, in the first region the kernel under consideration is limited by global memory bandwidth. In the second region, due to increasing data reuse, the limiting factor moves from global memory bandwidth to shared memory bandwidth\footnote{For some cases this might not occur because the shared memory capacity isn't large enough to store all data needed.}. This is the region in which the performance of the shared memory kernel is at its peak. The third and final region occurs when the number of taps is so high that it is no longer possible to fit enough data into memory to achieve good data reuse within shared memory and so performance suffers.

\begin{figure}[ht]
			\centering 
			\includegraphics[width=\linewidth]{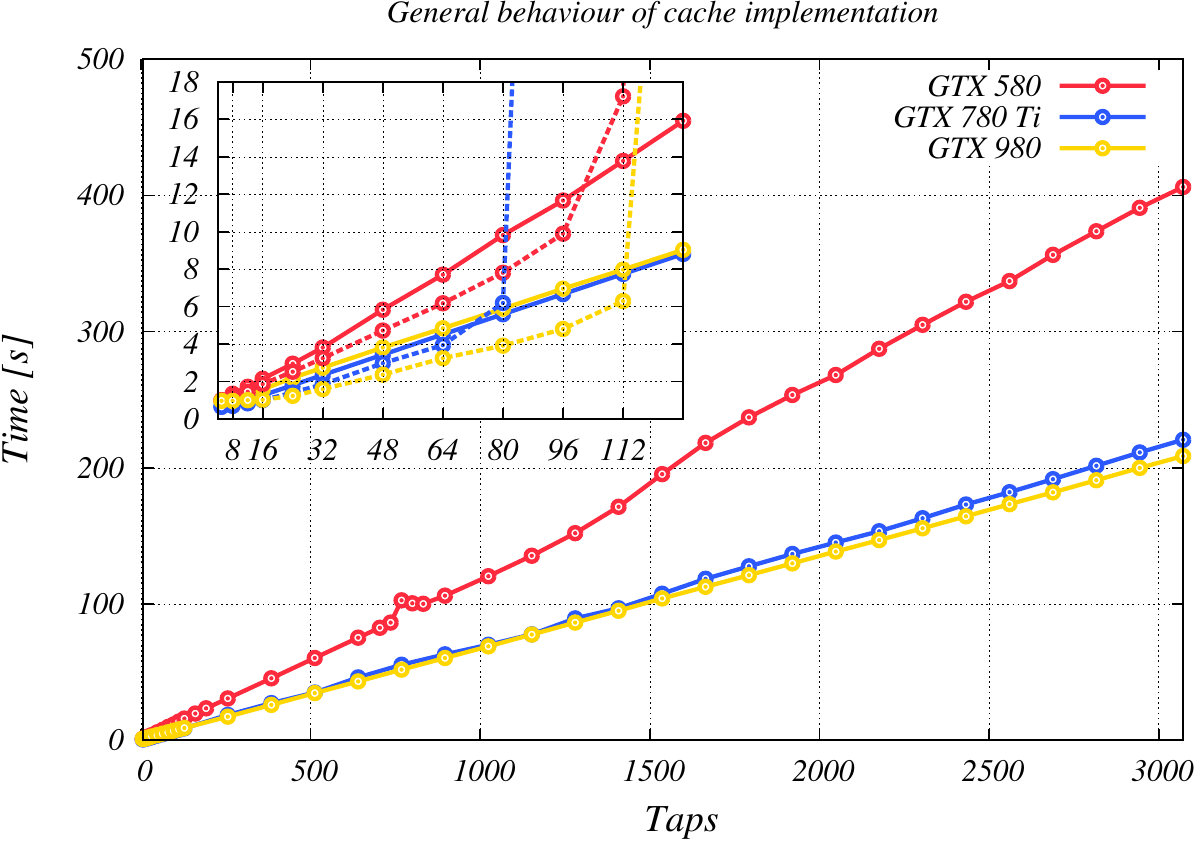}
			\caption{\label{fig:CKbehaviour} The general behaviour of our cache kernels for a very high number of taps and shared memory kernels for lower taps (dashed line in the inset plot). The execution time scales linearly with the number of taps in the case of our cache kernel.}
		\end{figure}

		Our cache kernels lack any constrains on the number of taps, thus they have essentially only two regions of distinct behaviour. The first where they are limited by global memory or L2 cache bandwidth and the second is where they are limited by L1 or texture cache bandwidth. The behaviour for a very large number of taps is shown in Figure~\ref{fig:CKbehaviour}.
		
	\subsection{GPU Behaviour}
		As generations of the \nv GPU evolve, so must the optimal implementation for them, making use of newly introduced features. The Kepler generation of GPUs enabled easy access to the texture cache (a read-only data cache), provided a larger and faster L2 cache along with a double data rate (8 byte transfer mode per bank) from shared memory. It also introduced shuffle instructions.
		
\begin{figure*}[ht]
		\begin{center}
			\includegraphics[width=.99\linewidth]{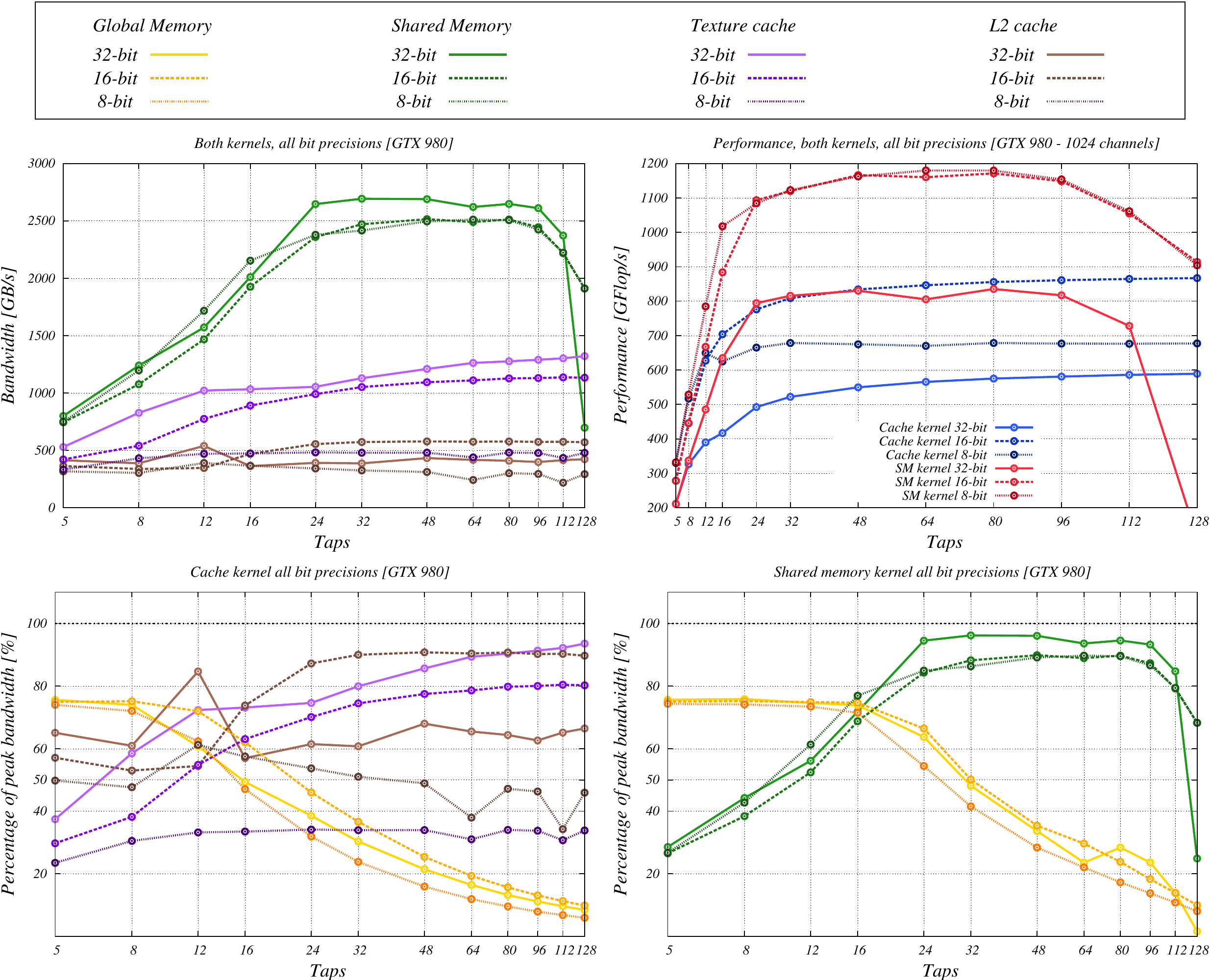}
			\caption{The compute performance and bandwidth for the GTX\,980 for our kernels. Bandwidth in absolute values (top left), performance in effective FLOPs (top right). Percentage of measured peak bandwidth for cache kernel (bottom left) and for our shared memory kernel (bottom right), it can clearly be seen where the 8-bit cache kernel hits the type conversion limit.}
			\label{fig:980all}
		\end{center}
		\end{figure*}

\begin{figure}[ht]
			\centering 
			\includegraphics[width=\linewidth]{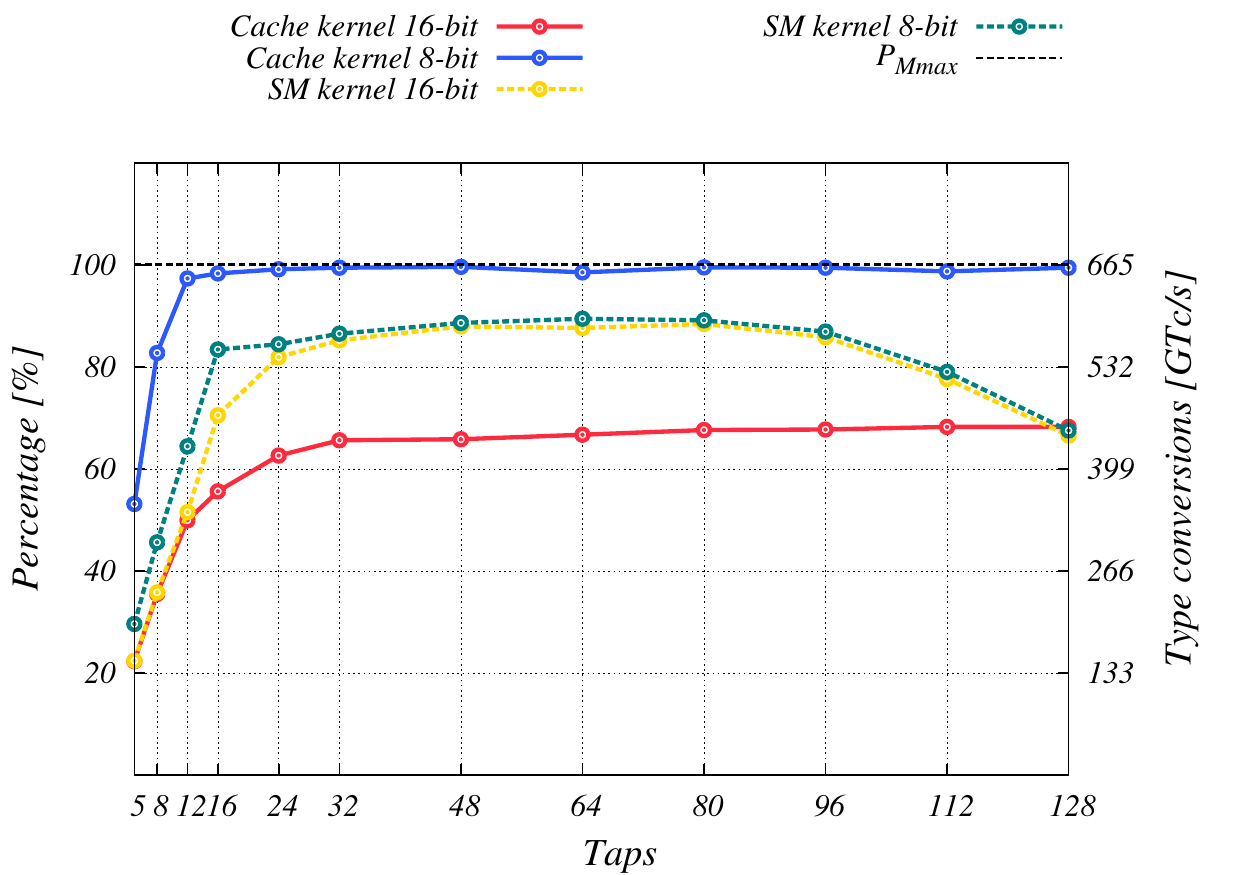}
			\caption{The utilisation of type conversions (TC) of GTX\,980 by both the cache (solid lines) and shared memory (dashed lines) kernels. The very high utilisation in the case of the 8-bit cache kernel (blue solid line) is due to superfluous type conversions. The decline in utilisation of type conversions in the case of the shared memory kernel is due to insufficient shared memory size and thus in-efficient data reuse for higher taps.}
			\label{fig:TC}
		\end{figure}

		The Maxwell generation of \nv ~GPU unified the L1 and texture caches and increased the amount of shared memory per streaming multiprocessor (SM). However the maximum available shared memory per thread block remained at 48\,kB. Thus the maximum number of taps our code is capable of processing remain unaffected. The Maxwell generation can perform less type conversions per \textit{warp} per SM per clock than the Kepler generation, about 60\% less. However despite this our Maxwell kernel achieves a higher peak performance in type conversions than our Kepler kernel. Figure~\ref{fig:980all} reports the performance and bandwidth as reported by \nv profiler. The type conversion utilisation for kernels with lower precision is shown in Figure~\ref{fig:TC}.

			\subsubsection{Cache kernel}
			The performance of our cache kernel is depicted in Figure~\ref{fig:980all}. Top left shows absolute values of bandwidth, bottom right fractions of measured or theoretical bandwidth values and top right performance. Our 32-bit cache kernel (straight line) is at first limited by global memory, then by L2 and finally by texture cache bandwidth, with utilisation up to 90\% of peak. There is a transition phase between 12 and 32 taps. The efficiency $\epsilon$ of our cache kernel is high, mostly above 90\%.
			
			The 16-bit implementation (dashed line) is bandwidth bound. Both L2 and texture cache achieve high fractions of peak measured bandwidth.

			The 8-bit implementation (dotted line) is more interesting. The most important limiting factor is the type conversion rate. The 8-bit kernel issues twice as many instructions as needed i.e. for each \texttt{int} to \texttt{float} instruction \texttt{I2F.F32.U16} there is one extra \texttt{int} to \texttt{int} instruction \texttt{I2I.S16.U8}. This is due to the unpacking of a vector of data (uchar2). These extra type conversions contribute significantly to the total amount of type conversions performed consequently performance is reduced. Our 8-bit cache kernel is slower than our 16-bit cache kernel. The utilisation of type conversion by our 8-bit cache kernel can be seen in figure \ref{fig:TC} as a solid blue line. The effect on performance can clearly be seen in Figure~\ref{fig:980all}. The 8-bit cache kernel becomes compute bound very quickly as it has above 80\% of measured peak at 8 taps and above 90\% at 12 taps.
			
			The Kepler generation exhibited similar behaviour as in the Maxwell 8-bit kernel case. The Kepler 8-bit kernel also produce twice as many type conversion instructions as needed, but for the Kepler generation these extra instructions do not seem to limit the performance. The 16-bit kernel performs the expected number of type conversions with high utilisation of the special function unit (SFU). The 8-bit kernel performs better while computing additional \texttt{int} to \texttt{int} instructions, suggesting that these additional conversions (\texttt{I2I}) are not computed by SFU and are not a limiting factor.
			
			The Fermi architecture did not show any deviations and produced the expected number of type conversions for all lower bit precision's.
		
			\subsubsection{Shared Memory Kernel}
			The behaviour of the shared memory kernel is depicted in Figure~\ref{fig:980all}. Top left shows absolute values of bandwidth, bottom right fractions of theoretical or measured peak bandwidth and top right performance. The shared memory kernel for 32-bit precision is at first limited by global memory. Global memory bandwidth remains almost unchanged up to the 16 taps, between 16 and 24 taps a transition occurs towards the limiting factor being shared memory bandwidth. At 24 taps shared memory bandwidth is already above 90\% of the theoretical peak. The peak in performance and in shared memory bandwidth is reached at 32 taps. A similar transition can be seen in both lower precision kernels. Beyond 32 taps the data reuse becomes less effective and the performance drops. Again lower precision kernels follow same pattern, but with a different number of taps giving the peak performance.

			Perhaps a deeper understanding can be gained by further considerations of mean data reuse $R_\up{T}$ and efficiency\footnote{to get $B_\up{W}$ and $B_\up{R}$ we have used the number of transactions reported by the \nv profiler} $\epsilon$ of our 32-bit implementation. The result together with achieved occupancy is shown in Figure~\ref{fig:meanreuse}. Mean data reuse $R_\up{T}$ remains relatively high, above 60\%, up to and including 32 taps. When considering efficiency $\epsilon$ it can be seen that any inefficiency in data reuse is compensated for by the cache\footnote{either texture, L2 or L1}. If this was not the case efficiency would be lower, since data would be loaded from global memory and thus increase $B_\up{R}$, hence lowering the efficiency $\epsilon$. A drop occurs at 64 taps, which should also be visible in performance, but is not. The resolution to this might be a high efficiency value $\epsilon$ which compensates for poor data reuse. Indeed the unified cache bandwidth utilisation can be observed to rise from $279\,\up{GB/s}$ at 48 taps to $328\,\up{GB/s}$ at 64 taps. At 80 taps it drops to $193\,\up{GB/s}$ which would also coincide with massive drop in efficiency $\epsilon$. When the number of taps is close to the maximum number of taps the data reuse is poor because the shared memory is not large enough to hold higher values of $S_\up{B}$. Thus most of the memory space is taken by columns which cannot be subsequently reused.
			
			The 8-bit shared memory kernel shows the same superfluous type conversions as its cache counter part. These are again due to the unpacking of a vector of data which is accessed via the \texttt{\_\_ldg()} intrinsic instruction. However while they are still present they only represent 2\% of total number of type conversion instructions executed by the code and thus the impact on performance is negligible. Loading data from shared memory, where the most memory requests are directed, does not cause extra type conversions since \texttt{I2F.F32.U8} is used directly. Both kernels have similar performance, this is due to the  kernels being compute bound (8-bit from 16 taps with 80\% of measured peak of TC, 16-bit from 24 taps with 90\% of measured peak of TC) which limits the maximum performance. Moreover, our 8-bit kernel loads data which fits into 32-bit words as does our 16-bit kernel (our 32-bit kernel requires a 64-bit word), thus similar configurations (with similar $R_{\mathrm{T}}$) are chosen for it. Both effects result in nearly identical behaviour for higher taps.

			In the case of our shared memory kernel we have also investigated other means of accessing memory or storing data. In an attempt to increase global memory bandwidth we have tried doubling the number of requested cache-lines,hence creating more memory requests in flight. This however greatly reduces the value for the maximum number of taps. Marginal performance benefit is observed when the kernel is bound by global memory bandwidth. In another variation we have tried to store data as 32-bit words, thus avoiding costly bit-conversions. This did not provide any performance boost because the code was limited by shared memory bandwidth for a smaller number of taps.

\begin{figure}[ht]
				\centering 
				\includegraphics[width=\linewidth]{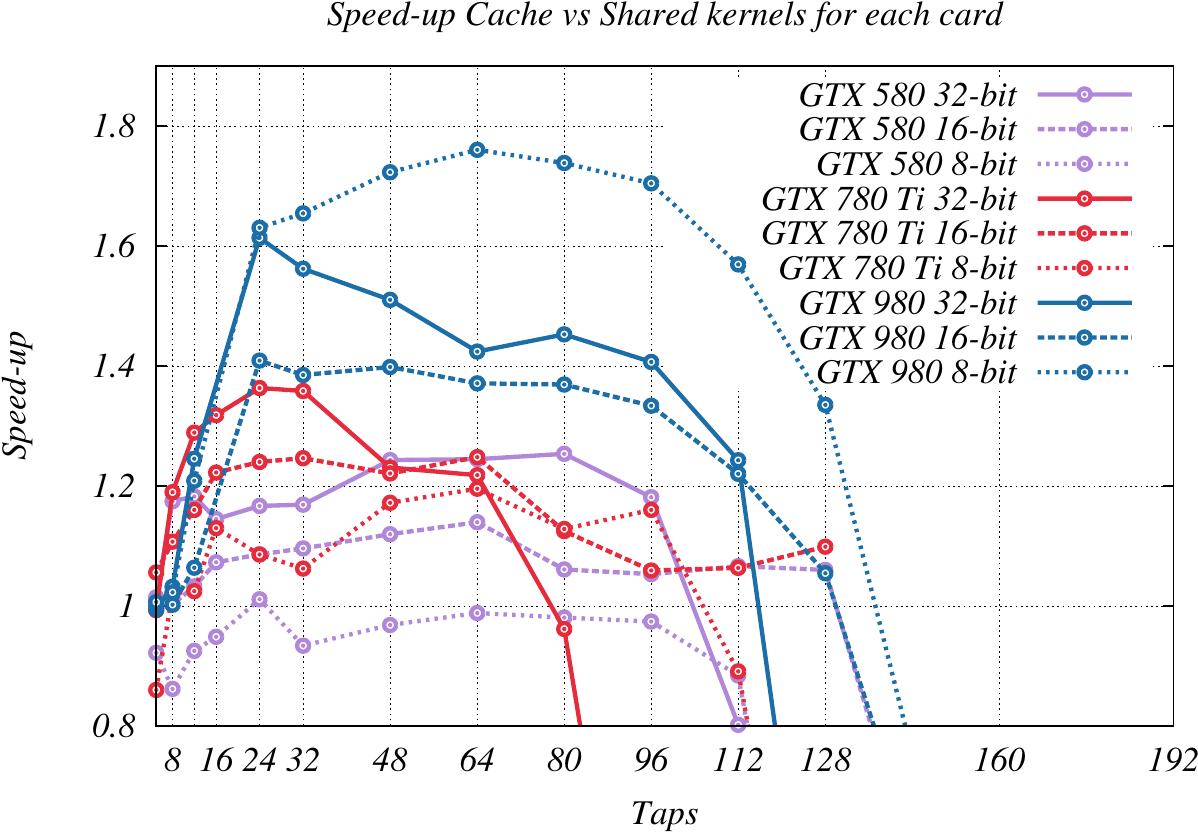}
				\caption{\label{fig:Speeduptaps} The relative speedup (Cache vs. shared memory) and its dependence on the number of taps for all bit precision's and all generations.}
			\end{figure}

\begin{figure}[ht]
				\centering 
				\includegraphics[width=\linewidth]{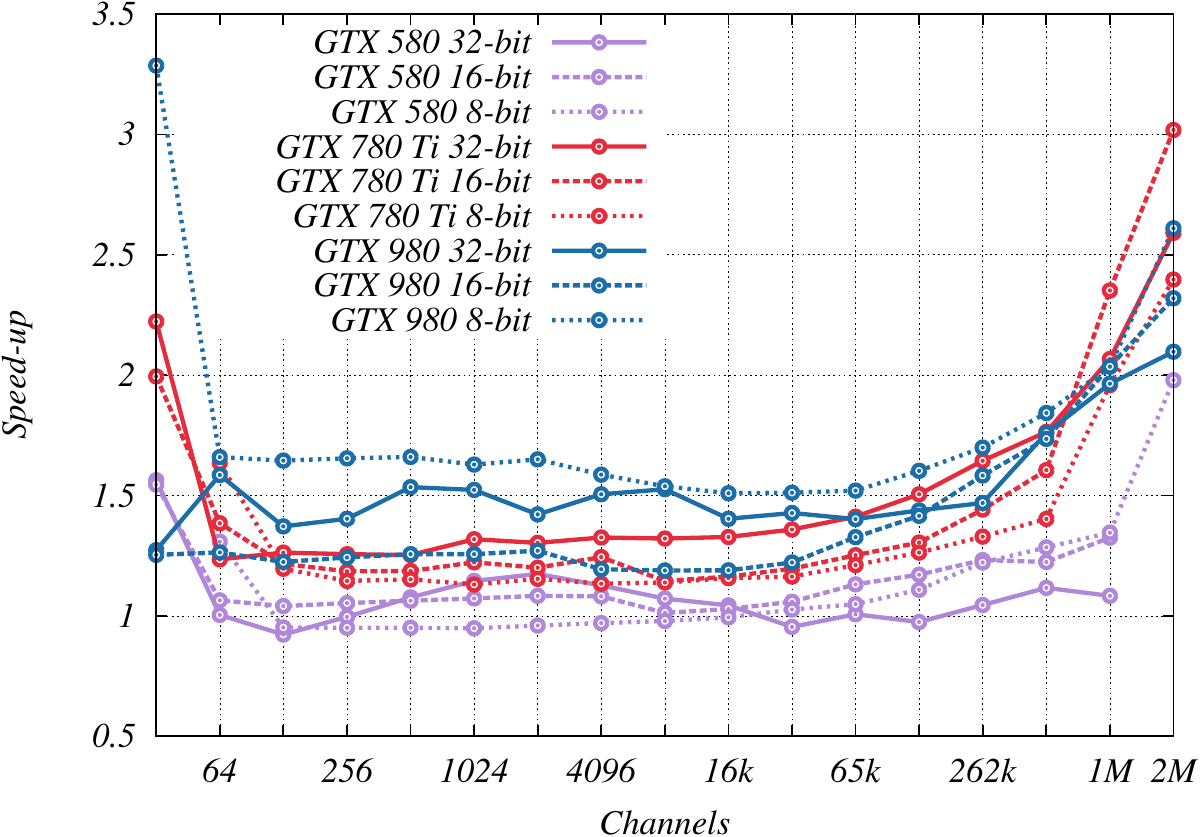}
				\caption{\label{fig:Speedupchannels} The relative speedup (Cache vs. shared memory) and its dependence on the number of channels for all bit precision's and all generations. }
			\end{figure}

			Lastly we compare the relative speedup of our shared memory kernels compared to their cache based counterparts for all bit implementations. Results are presented in Figures~\ref{fig:Speeduptaps} for different numbers of taps (channel number fixed to 1024) and \ref{fig:Speedupchannels} for different channels (taps number fixed to 16). The benefit of using the shared memory implementation kernel can be seen especially on Figure~\ref{fig:Speeduptaps}, e.g. for Fermi: around 10--20\%, Kepler: 15--20\% and Maxwell: 30--70\%.

	\subsection{Xeon Phi}
\begin{figure*}[ht]
				\centering 
					\includegraphics[width=\linewidth]{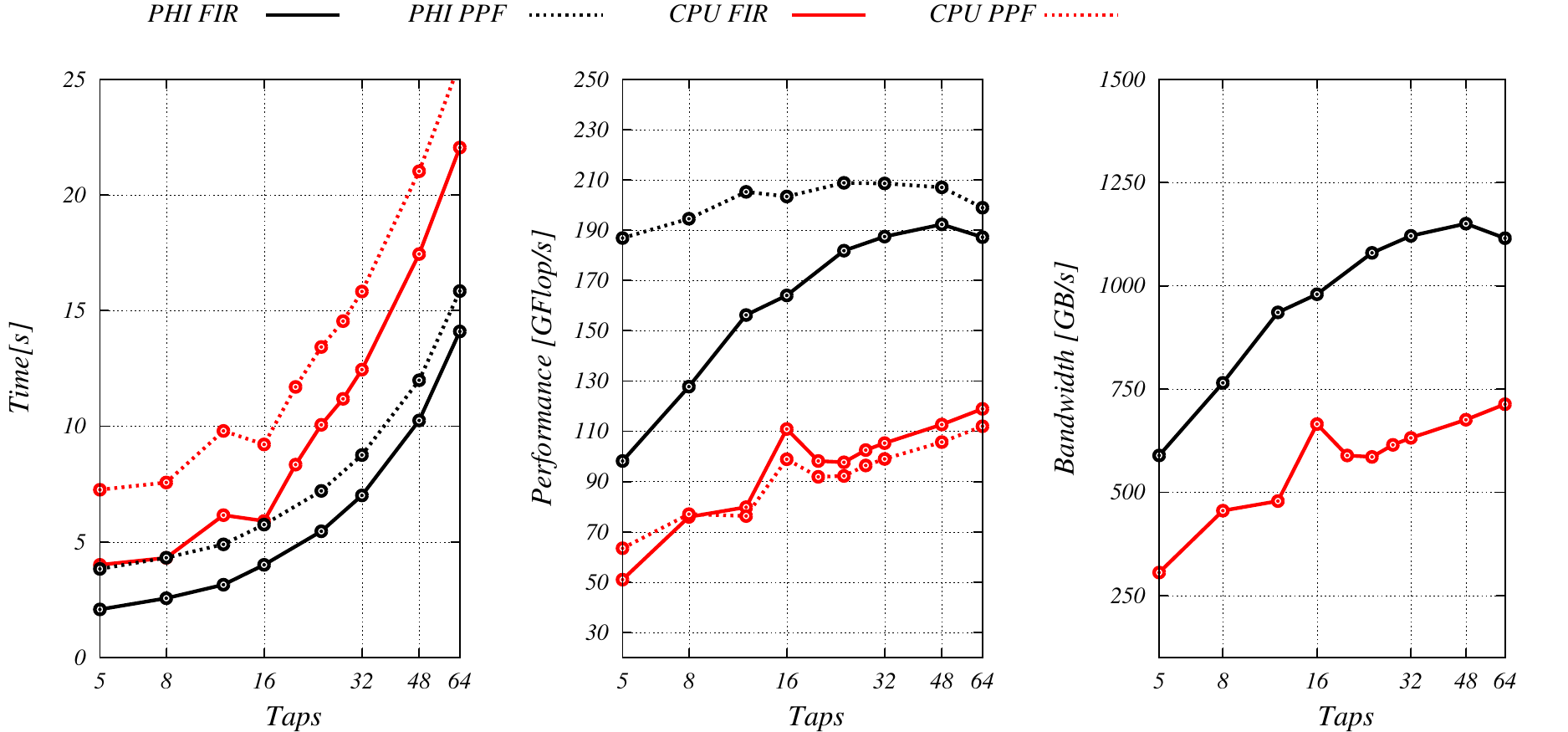}
				\caption{\label{fig:Phi_t} The execution time (left), effective FLOPs (middle) and computed L1 cache bandwidth (right) for Xeon Phi and CPU implementations of the PPF as a function of the number of taps.}
			\end{figure*}

\begin{figure*}[ht]
				\centering 
					\includegraphics[width=\linewidth]{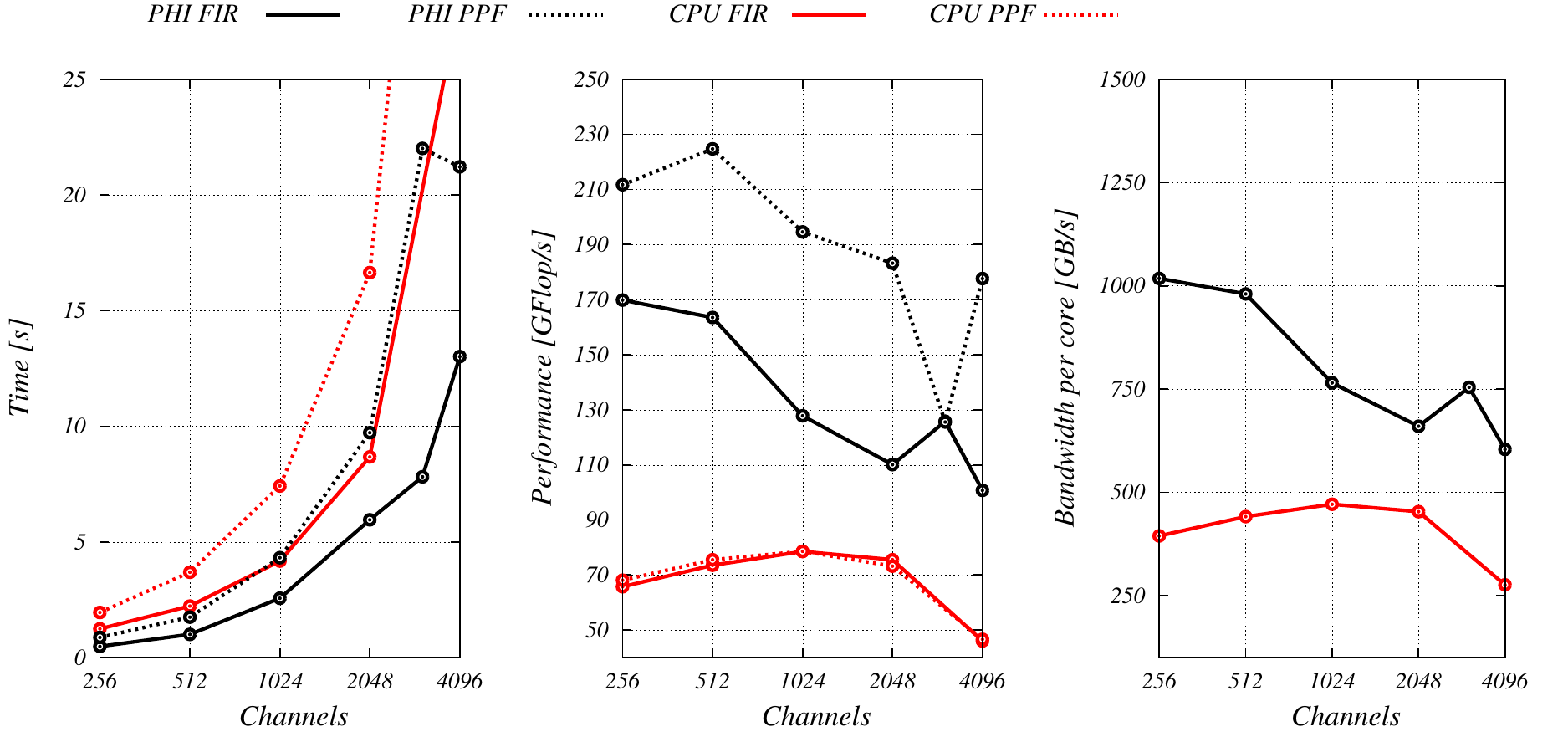}
				\caption{\label{fig:Phi_c} The execution time (left), performance (middle) and computed L1 cache bandwidth (right) for Xeon Phi and CPU implementations of the PPF as a function of the number of channels. The peak in execution time or drop in performance for the complete PPF at 3072 channels is caused by poor performance of the FFT library for a number of channels which are not a power of 2.}
			\end{figure*}

		The PPF is bandwidth bound by low level cache on both the Xeon Phi and dual CPU platforms. Thus the maximum achievable speedup between the two platforms will be governed by the behaviour of these caches. Our achieved speedups vary from $1.47\times$ to $1.95\times$. Results can be seen in Figure~\ref{fig:Phi_t} and \ref{fig:Phi_c}. In general both code's behaviour is comparable to that of the GPU's. At first the code is limited by global memory, but with an increasing number of taps and thus data reuse, the limitation shifts to cache bandwidth. Our calculated L1 bandwidth is showed in Figures~\ref{fig:Phi_t} and \ref{fig:Phi_c}. We do not have 16-bit and 8-bit results since intrinsic instructions for manipulating lower precision data are not supported the current generation of Xeon Phi (Knights Corner).
			
\section{Results}
\label{sec:Results}
	In this section we present our findings, then compare our results to the results of others and finally present results for the whole PPF for a large set of channels and taps.
	
	\subsection{Data transfers via PCIe}
\begin{figure*}[ht]
			\centering 
			\begin{tabular}{c}
			\includegraphics[width=.8\linewidth]{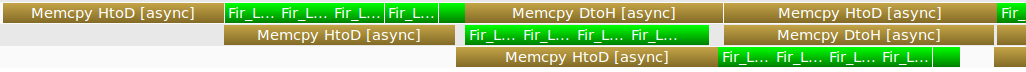} \\
			\textit{Latency hidden}\\
			\includegraphics[width=.8\linewidth]{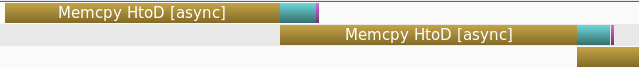} \\
			\textit{Latency not fully hidden} \\
			\end{tabular}
			\caption{An example of latency hiding behind computations as showed by the \nv profiler. Fully hidden computation (top), which is possible only when the card has two copy engines. The case when a card can transfer data only in one direction and latency cannot be fully hidden (bottom). }
			\label{fig:streams}
		\end{figure*}

		An important characteristic of current accelerator devices is the ability to hide computation behind data transfers. If the time spent on calculation is equal to (or less than) the time needed for data transfer between the host computer and the accelerator device then the limiting factor of the data processing is the PCIe bandwidth. In the case of GPU's bi-directional transfer can be achieved using asynchronous copy engines. If a card has only one such copy engine we would not be able to hide all computations. With two copy engines we can effectively hide the computations completely, this is demonstrated in Figure \ref{fig:streams}. The Xeon Phi has also ability to hide computation behind data transfers, however we could not achieve full bi-directional transfers with the Xeon Phi hardware and software that we employed for these tests\footnote{We believe this was due to a software bug in Intel software.}.
		
	\subsection{Comparison to others}
		We compare our implementation of the polyphase filter with two other implementations. We have made every effort to produce fair comparisons, however direct comparison has proven to be difficult. We first compare the implementation of a polyphase filter by J. Chennamangalam et al. \cite{PAS:9452994} created for the VEGAS spectrometer. This implementation has emphasis on a high number of channels. Direct straight forward comparison is not possible, since the VEGAS implementation uses two complex polarisation's in 8-bit precision per frequency channel. We consider only one complex polarisation in our input data. To overcome this difference it is sufficient to double number of channels in our input data. So if we are to calculate $100$ channels with the VEGAS implementation we must set the number of channels we compute to $200$ in order to process same amount of data. Our implementation in this scenario has the disadvantage of having to use two coefficients, whereas VEGAS uses only one. 
		
		Moreover VEGAS uses so called sub-bands which are essentially data streams from different detectors. Our input data format does not take sub-bands into consideration. There are two ways in which we could adopt this data format within our code. The first is to process data from different detectors in packets. This configuration however requires some data rearrangement or other changes thus we do not consider it for comparison. The second is to consider sub-bands as part of a single spectra. As such a sub-band is basically a coordinate within a spectra which signifies where the sub-band starts. Our comparisons to VEGAS use 65536 channels because this is the maximum number of channels VEGAS can process with 64 taps. For a lower number of taps we run VEGAS with as many sub-bands as possible to ensure best performance. The number of channels for our code is calculated as $C_\mathrm{our}=2C_\mathrm{V}N_\mathrm{Sub}$, where $C_\mathrm{V}$ is the number of channels used for VEGAS and $N_\mathrm{Sub}$ is the number of sub-bands. The results of our comparisons can be seen in the Table~\ref{tab:jayanth}.
		
		\begin{table*}[ht]
\centering
\caption{shows comparison of PPF code for VEGAS spectrometer and our PPF code. PPF in this case is dominated by FFT due to data format. Our better performing FIR filter part thus shows only for higher number of taps where execution time of FIR and FFT are equalized.}
\begin{tabular}{l l *{4}{r r}}
\toprule
\multirow{2}{*}{\hphantom{1}8 bit} & 
\multirow{2}{*}{Kernel} &
\multicolumn{8}{c}{Execution time (Speed-up) for different taps} \\
\cmidrule(r){3-10}
&
& 
\multicolumn{2}{c}{8 taps}   & 
\multicolumn{2}{c}{16 taps}  & 
\multicolumn{2}{c}{32 taps}  & 
\multicolumn{2}{c}{64 taps}   \\
\midrule
\multirow{3}{*}{GTX 580} & VEGAS & 425.030 & & 379.154 & & 367.954 & & 529.319 &\\
& Cache &  425.386 & (0.99) & 345.279 & (1.10) & 260.624 & (1.41) & 276.015 & (1.92) \\
& SM &  412.141 & (1.03) & 331.596 & (1.14) & 245.082 & (1.50) & 274.586 & (1.93) \\
\cmidrule(r){3-10}
\multirow{3}{*}{GTX 780} & VEGAS & 217.528 & & 206.206 & & 227.145 & & 334.483 &\\
& Cache & 211.970 & (1.03) & 176.757 & (1.17) & 159.355 & (1.43) & 195.509 & (1.71) \\
& SM & 195.009 & (1.12) & 160.772 & (1.28) & 137.890 & (1.65) & 180.143 & (1.86) \\
\cmidrule(r){3-10}
\multirow{3}{*}{GTX 980} & VEGAS & 244.503 & & 247.199 & & 313.864 & & 472.373 &\\
& Cache & 222.178 & (1.10) & 186.960 & (1.32) & 186.217 & (1.69) & 216.270 & (2.18) \\
& SM & 205.995 & (1.19) & 162.472 & (1.52) & 154.799 & (2.03) & 169.727 & (2.78)
 \\
\bottomrule
\end{tabular}
\label{tab:jayanth}
\end{table*}
		
		The VEGAS implementation uses a specific data structure which is demanding on FFT execution time. Since FFTs for a smaller number of taps dominate the execution time of the PPF, our faster running FIR becomes less visible. For a higher number of taps, where the FIR filter execution time becomes comparable the FFT's, our PPF is faster.
		
		The second code we have compared to is a code developed for the LOFAR radio telescope by Karel van der Veld \cite{vanderVeldt:2012:PFG:2286976.2286986}. The code is very specific and it is written with the aim of exploiting registers. We find that the code has register spills and these limit the number of channels that can be processed, the maximum being $256$, this being the main reason why this code scales poorly. The number of taps can be set to 4, 8, 16, 32 and 64. From the description in \cite{vanderVeldt:2012:PFG:2286976.2286986}, we conclude that the best configuration is for 256 channels and 16 taps. The results of our comparison to the LOFAR code is presented in Table~\ref{tab:LOFAR}. To produce Table~\ref{tab:LOFAR} we have used timings from the \nv profiler. The LOFAR implementation uses CPU clocks to measure performance, which we found to be unreliable if the CPU changes clock frequency during execution. We also note that our code is much more versatile in terms of both channels and taps and has far better scaling with the number of channels and taps.
		
		\begin{table*}[ht]
\centering
\caption{shows comparison of code for LOFAR telescope and our code. Poor performance of LOFAR code on GTX 580 for high taps value is mostly due to register spill. This is no longer present with following generations. Our code is always faster then LOFAR code. Direct comparison can be performed up to 16 taps. After that LOFAR code seems to be reusing coefficients, where we do not. The  coefficients in our implementation can be arbitrary.}
\begin{tabular}{l l *{5}{r r}}
\toprule
\multirow{2}{*}{\hphantom{1}8 bit} & 
\multirow{2}{*}{Kernel} &
\multicolumn{10}{c}{Execution time (Speed-up) for different taps} \\
\cmidrule(r){3-12}
&
& 
\multicolumn{2}{c}{4 taps} &   
\multicolumn{2}{c}{8 taps} &   
\multicolumn{2}{c}{16 taps} &  
\multicolumn{2}{c}{32 taps} &   
\multicolumn{2}{c}{64 taps}   \\
\midrule
\multirow{3}{*}{GTX 580} & Lofar & 58.077 & & 60.141 & & 63.663 & & 649.901 & & 3\,450.253 &\\
& L1 & 26.327 & (2.21) & 28.647 & (2.10) & 38.587 & (1.65) & 58.695 & (11.07) & 99.027 & (34.84) \\
& SM & 26.625 & (2.18) & 30.792 & (1.95) & 39.955 & (1.59) & 61.713 & (10.53) & 100.042 & (34.49) \\
\cmidrule(r){3-12}
\multirow{3}{*}{GTX 780 Ti} & Lofar & 34.951 & & 39.052 & & 40.311 & & 65.735 & & 162.130 &\\
& L1 & 15.896 & (2.20) & 18.731 & (2.08) & 26.965 & (1.49) & 42.584 & (1.54) & 74.779 & (2.17) \\
& SM & 16.266 & (2.15) & 19.037 & (2.05) & 25.045 & (1.61) & 40.416 & (1.63) & 63.267 & (2.56) \\
\cmidrule(r){3-12}
\multirow{3}{*}{GTX 980} & Lofar & 55.678 & & 57.038 & & 58.123 & & 61.316 & & 163.423 &\\
& L1 & 26.081 & (2.13) & 26.256 & (2.17) & 33.752 & (1.72) & 51.027 & (1.20) & 85.751 & (1.91) \\
& SM & 26.338 & (2.11) & 26.855 & (2.12) & 31.303 & (1.86) & 43.180 & (1.42) & 77.709 & (2.10) \\
\bottomrule
\multirow{2}{*}{16 bit} & 
\multirow{2}{*}{\hphantom{Kernel}} &
\multicolumn{10}{c}{} \\
\cmidrule(r){3-12}
&
& 
\multicolumn{2}{c}{4 taps} &   
\multicolumn{2}{c}{8 taps} &   
\multicolumn{2}{c}{16 taps} &  
\multicolumn{2}{c}{32 taps} &   
\multicolumn{2}{c}{64 taps}   \\
\midrule
\multirow{3}{*}{GTX 580} & Lofar & 60.352 & & 62.274 & & 65.397 & & 658.564 & & 3\,455.297 &\\
& L1 & 28.689 & (2.10) & 32.128 & (1.94) & 44.595 & (1.47) & 70.014 & (9.41) & 121.293 & (28.49) \\
& SM & 28.250 & (2.14) & 32.261 & (1.93) & 42.942 & (1.52) & 65.586 & (10.04) &109.201 & (31.64) \\
\cmidrule(r){3-12}
\multirow{3}{*}{GTX 780 Ti} & Lofar & 36.925 & & 40.791 & & 41.317 & & 65.938 & & 163.124 &\\
& L1 & 17.027 & (2.17) & 19.448 & (2.10) & 27.055 & (1.53) & 52.691 & (1.25) & 73.285 & (2.23) \\
& SM & 18.083 & (2.04) & 19.479 & (2.09) & 25.408 & (1.63) & 38.926 & (1.69) & 62.584 & (2.61) \\
\cmidrule(r){3-12}
\multirow{3}{*}{GTX 980} & Lofar & 59.021 & & 59.927 & & 60.368 & & 63.433 & & 169.533 &\\
& L1 & 27.656 & (2.13) & 28.200 & (2.13) & 32.056 & (1.88) & 44.656 & (1.42) & 71.153 & (2.38) \\
& SM & 27.661 & (2.13) & 28.143 & (2.13) & 30.476 & (1.98) & 42.883 & (1.48) & 76.105 & (2.23) \\
\bottomrule
\label{tab:LOFAR}
\end{tabular}
\end{table*}

		\subsection{Sample rates}
\begin{figure*}[htp]
    \begin{minipage}[t]{.49\textwidth}
        \centering
        \includegraphics[width=\linewidth]{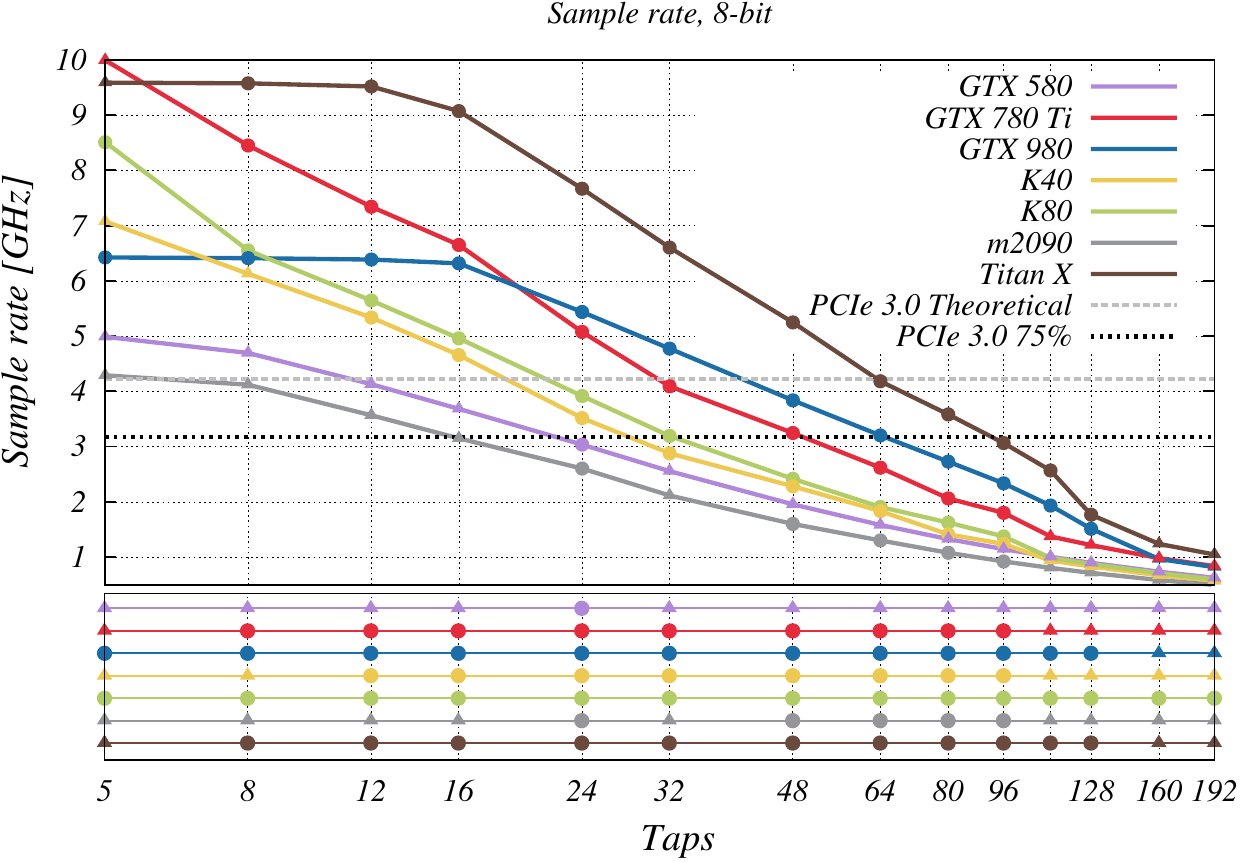}
    \end{minipage}%
    	\hfill%
    \begin{minipage}[t]{.49\textwidth}
        \centering
        \includegraphics[width=\linewidth]{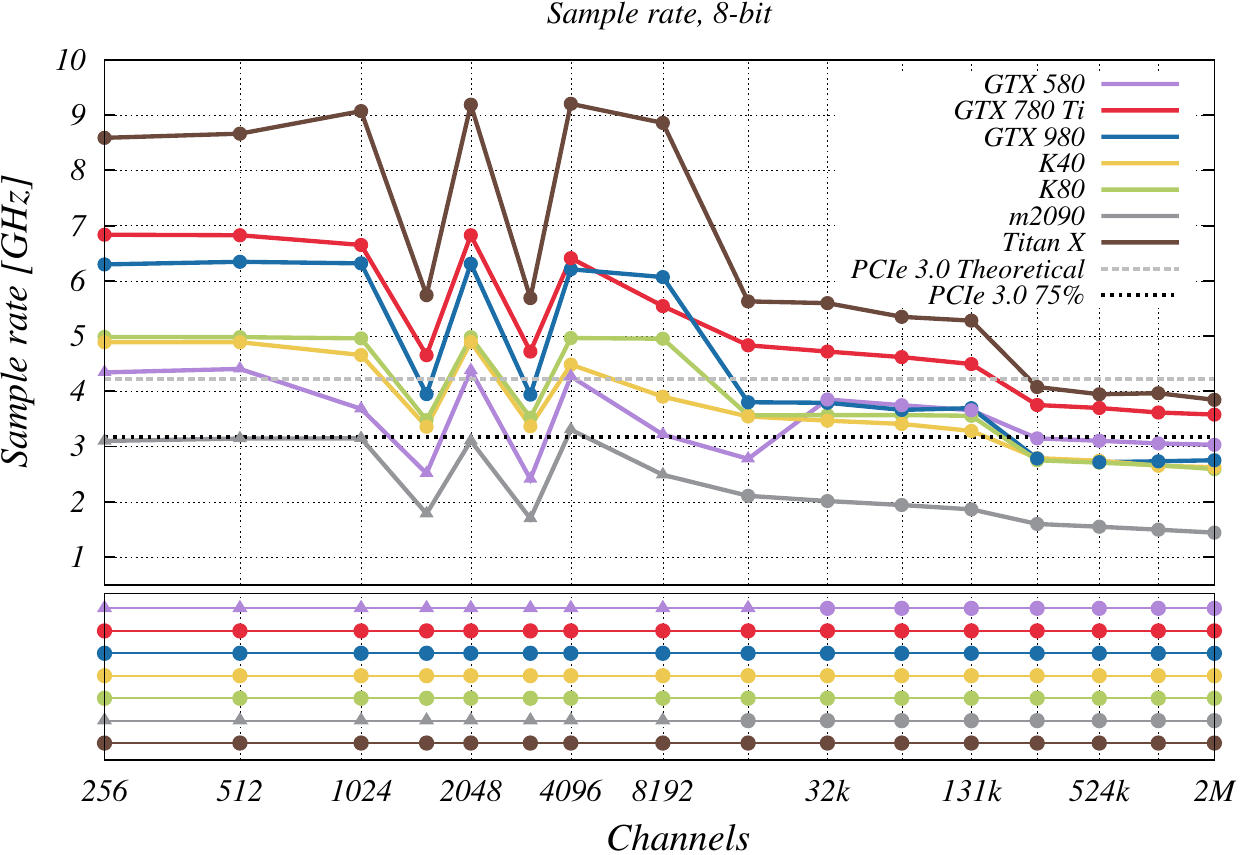}
    \end{minipage}\\
    \begin{minipage}[t]{.49\textwidth}
        \centering
        \includegraphics[width=\linewidth]{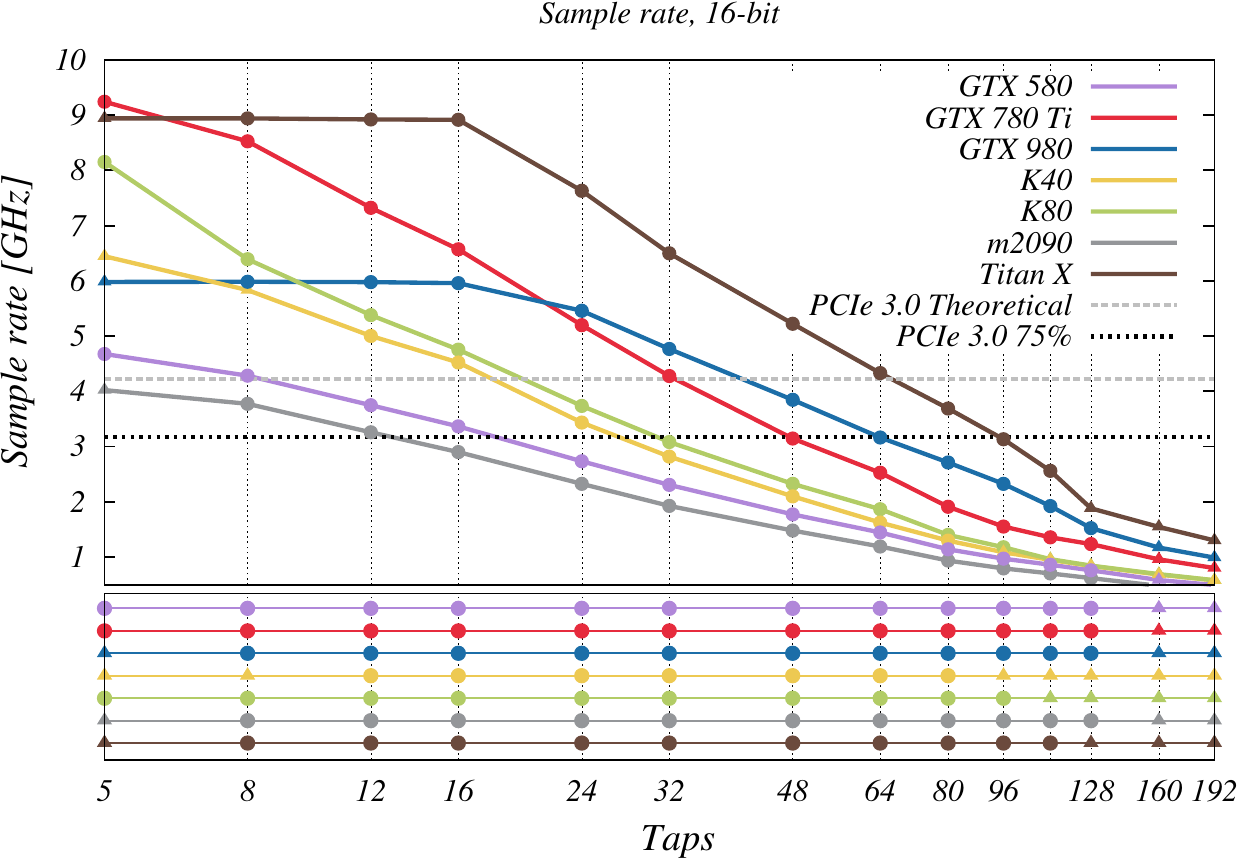}
    \end{minipage}%
    	\hfill%
    \begin{minipage}[t]{.49\textwidth}
        \centering
        \includegraphics[width=\linewidth]{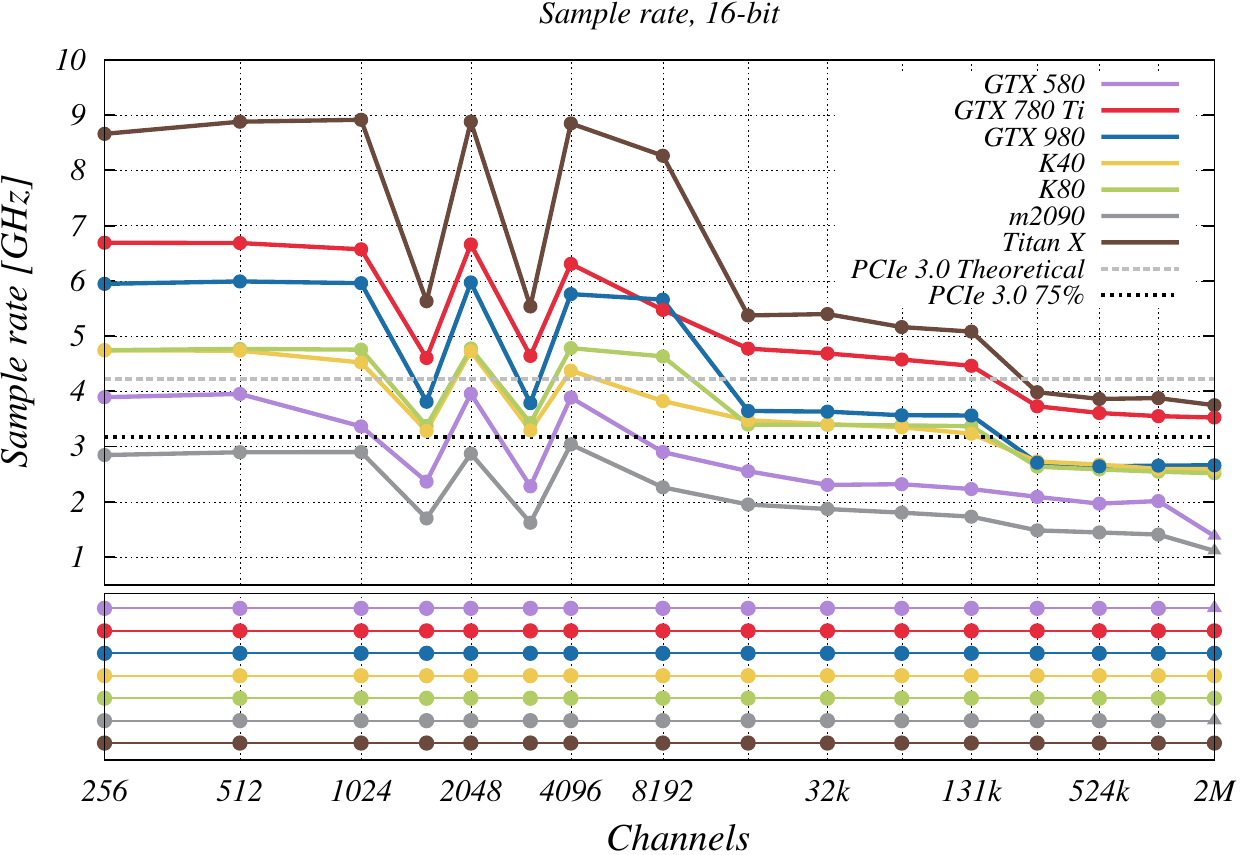}
    \end{minipage}\\
    \begin{minipage}[t]{.49\textwidth}
        \centering
        \includegraphics[width=\linewidth]{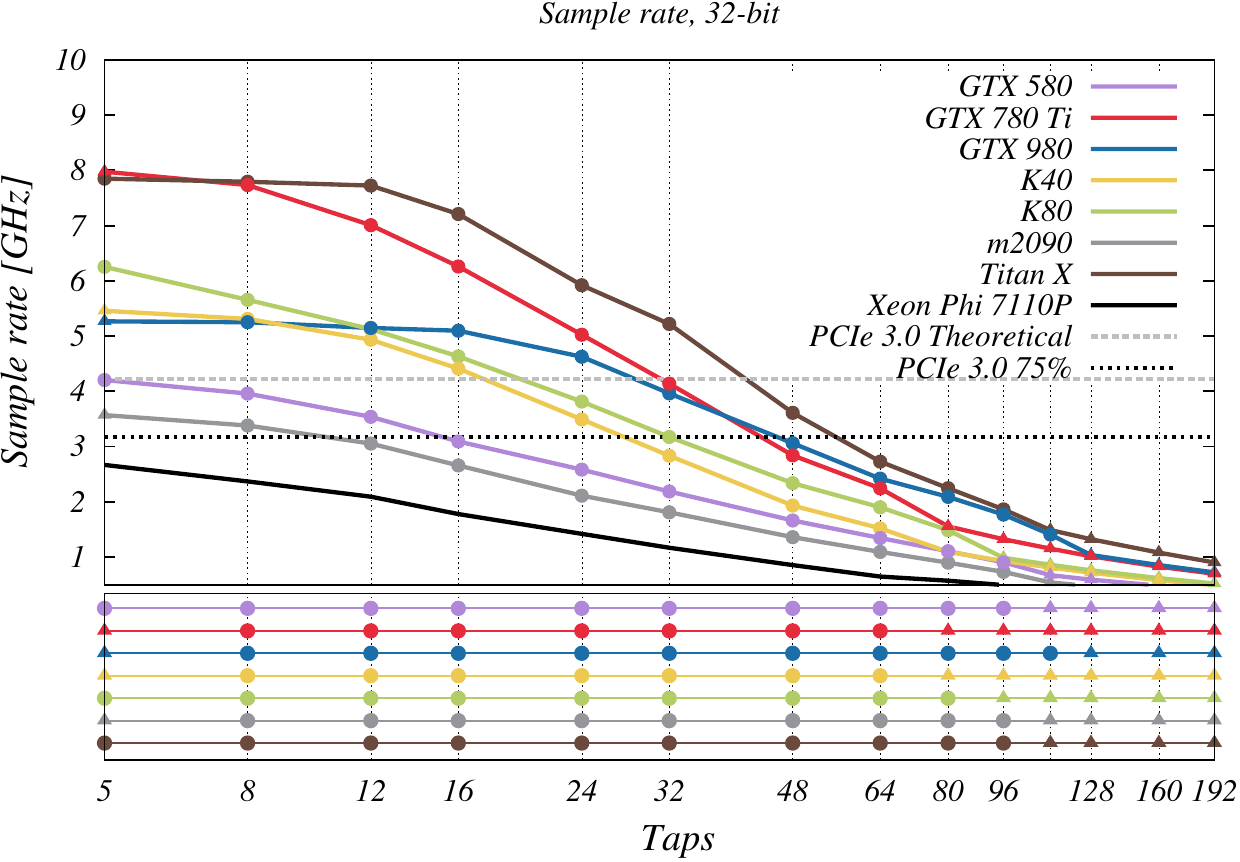}
    \end{minipage}%
    	\hfill%
    \begin{minipage}[t]{.49\textwidth}
        \centering
        \includegraphics[width=\linewidth]{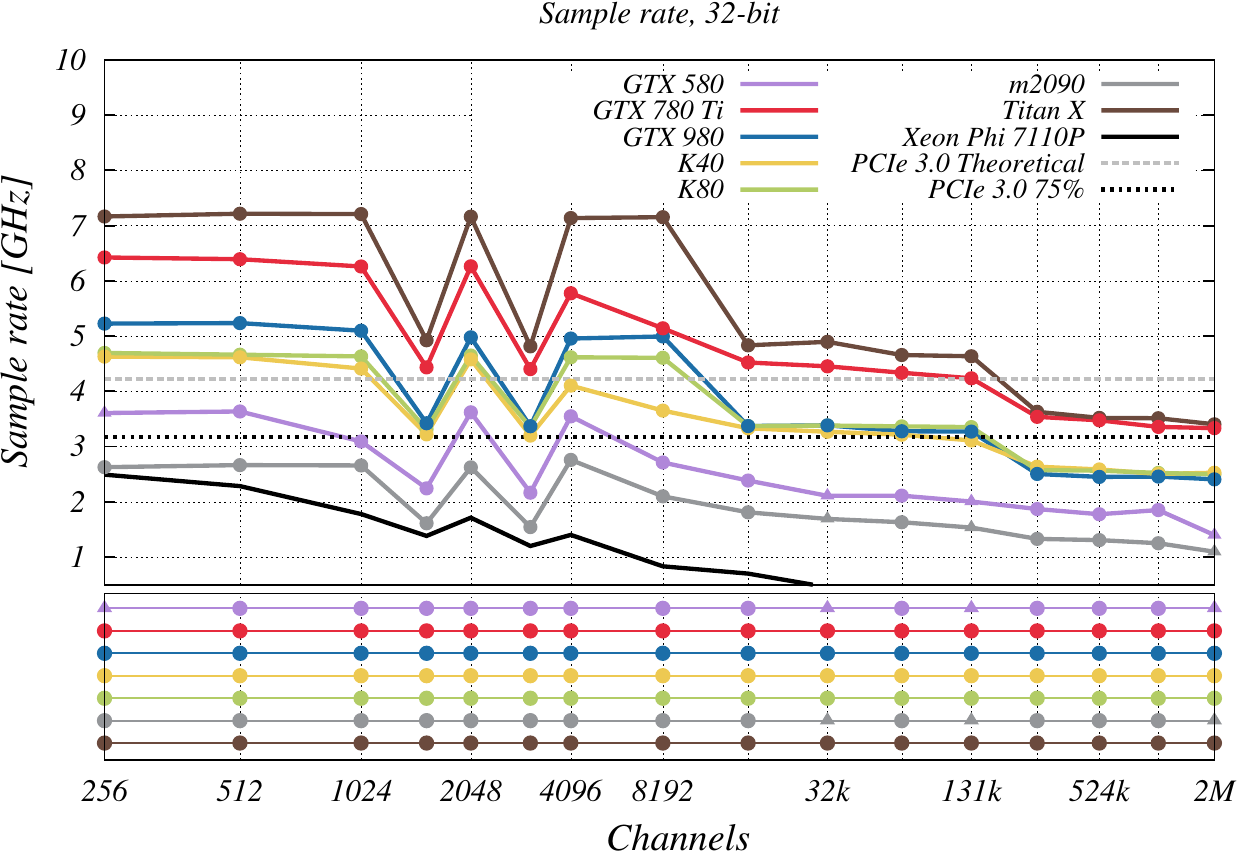}
    \end{minipage}
%
			\caption{Results of our PPF for a wide range of taps and channels. Results displayed are from the most optimal implementation for any given parameterisation. Triangles represent our cache based implementation, dots represent our shared memory implementation. All graphs show a horizontal line which describes the PCIe 3.0 bus bandwidth in terms of sample rate. These lines are for the theoretical value (grey dashed) and 75\% of the theoretical value which is an estimate of what could be realistically achieved. Sample rates above these lines indicate that a GPU is able to process all of the data we can transfer to the card per second i.e. data are processed in real time or quicker. This also means that there will be room for other signal processing steps as well as the polyphase filter. Sharp drops in sample rates in the channel plots are caused by the FFT libraries poor performance for channel numbers which are not a power of 2. Some cards do not have results for all channels because they do not have enough memory to compute them. Note that the K80 has two physical GPUs, but our results use only one of these.}
			\label{fig:SRAll}
		\end{figure*}

		We have chosen to present results of our PPF implementation in terms of the number of samples a device can process per second, i.e. the maximum sample rate that can be processed. The maximum sample rate allows us to present real-time performance without limiting ourselves to the specifics of any one radio telescope. The performance of our PPF for all bit precision's is shown in Figure~\ref{fig:SRAll}. 

		\begin{figure}[ht]
			\centering 
			\includegraphics[width=\linewidth]{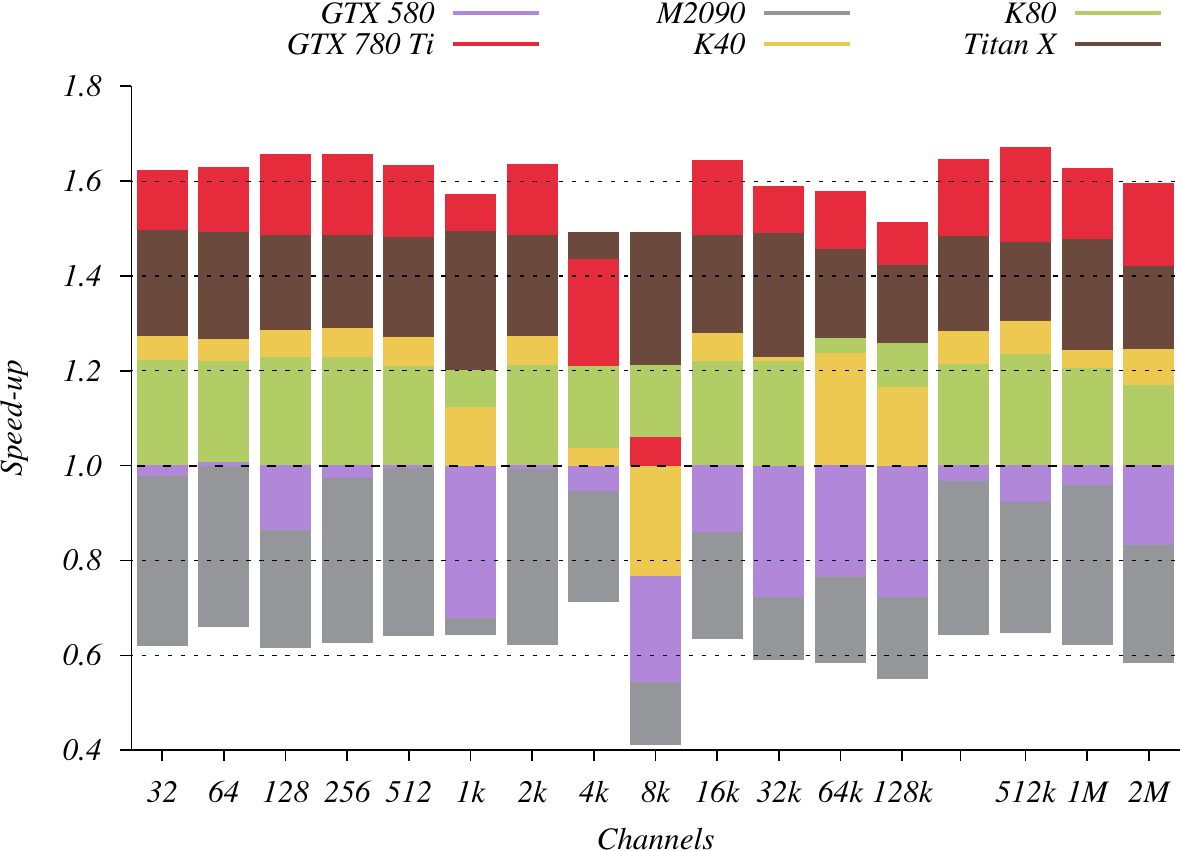}
			\caption{ The speedup in FFT execution time on different cards with respect to GTX\,980. We see that the FFT execution time on GTX\,980 takes longer than cards from the Kepler generation. The GTX\,980 execution time is sometimes comparable with the FFT execution time on the GTX\,580 (Fermi generation). Versions of CUDA used for these tests are those that are used throughout this paper, that is CUDA 6.5 for Fermi and Kepler generations and CUDA 7.0 for the Maxwell generation. We point out that the GTX\,780\,Ti, an older card, using an older CUDA version outperforms the newest card of the Maxwell generation (TITAN\,X) using a newer CUDA version 7.0.}
			\label{fig:FFTperf}
		\end{figure}
		
		Figure~\ref{fig:SRAll} shows several interesting behaviours. Firstly the GTX\,780\,Ti is sometimes able to outperform the latest flagship GPU, the TITAN\,X, and is better performing than the GTX\,980 below 24 taps due to the greater global memory bandwidth of the GTX\,780\,Ti. Also the FFT performance on the Maxwell architecture is generally slower than on the Kepler architecture, so even with a faster executing FIR filter on the Maxwell architecture, the slower FFT dominates performance and so the Maxwell architecture is disadvantaged. Our measured FFT performance on all of our tested cards is shown in Figure~\ref{fig:FFTperf}.  
		
		The plots on the right of Figure~\ref{fig:SRAll} give sample rates per second for a varying number of channels in the input data. We see that (prominently on the Maxwell generation) the variation of sample rate as a function of channels is small. The drops in performance are due to the FFT library performance, typically occurring when the number of channels are not a power of 2. In the worst cases the performance drops by about $2\times$ ($2.5\times$ for GTX\,580).
		
		The plots on the left side of Figure~\ref{fig:SRAll} give sample rates per second for a varying number of taps. Increasing the number of taps, increases the processing complexity per data element, thus the sample rate per second decreases with increasing taps. The GTX\,980 performance (for a small number of taps) is comparable with the performance of the GTX\,580, this is because for a low number of taps the performance is limited by global memory bandwidth which is comparable in case of these two cards. We can again see that up to 16 taps the GTX\,980 is able to sustain roughly same sample rate per second due to the PPF being global memory bandwidth bound, i.e. the global memory is unable to supply data quickly enough thus the shared memory bandwidth is not utilised to its full potential. This can also be seen in Figure~\ref{fig:980all}.
				
		We find that the GPUs tested from the Fermi generation, the GTX\,580 and M2090 both perform well when compared to the latest generation of GPUs, it is also worth noting that the performance of the GTX\,580 for lower bit precision is comparable to the K40.
		
		The Kepler generation offers, performance wise, more interesting choices. From three Kepler generation cards, two were scientific ones the K40 and K80 and on a gaming card the \mbox{GTX\,780\,Ti}. The GTX\,780\,Ti performs well for our application, being the second best performing card of all tested in some cases. This is due to the higher processing core and memory frequencies along with better performing FFTs, but from a real-time data processing perspective it is hindered by not having two copy engines. Both scientific cards, the K40 and K80 have two copy engines, however the K80 is a dual GPU card, i.e. two GPUs share the same PCIe bus. In this case the second GPU can be a burden or a benefit depending on whenever a real-time pipeline is bandwidth bound or compute bound respectively. The discrepancy between the theoretical peak type conversions performance and our measurements in the case of the Kepler generation is very high. We achieve only 22\% of the theoretical peak. However none of the tests that we have performed would suggest that a higher performance could be reached.
		
		As mentioned earlier the newest GPU generation (Maxwell) appears to issue superfluous instructions when performing type conversions on data that is transferred by using the \textit{\_\_ldg()} intrinsic instruction. Despite this, we find that the Maxwell generation can perform 50\% more type conversions than the Kepler generation, and more than twice as many type conversions as the Fermi generation. One however has to use a shared memory implementation in order not to be limited by the extra type conversion instructions generated by the texture cache. All of the Maxwell generation cards that we investigated have two copy engines and as such are well suited to real-time data processing. The fastest card, the TITAN\,X, is from the Maxwell generation. When not limited by global memory bandwidth, it outperforms the GTX\,780\,Ti by almost 30\% in some cases.

	
\section{Conclusions}
\label{sec:Conclusions}
	We have implemented a polyphase filter on three platforms GPU, Xeon Phi and Xeon CPU. Our GPU implementations have three bit-precision versions, 32-bit, 16-bit and 8-bit. Our codes place few restrictions on the number of channels or taps that can be processed and we do not exploit any symmetries that FIR filter coefficients might have. We have demonstrated that our implementation is faster than the other two published implementations we know of. We have presented results in the context of real-time data processing pipelines and discussed associated hardware benefits and deficiencies. For real-time data processing on GPUs one has to use scientific cards in the case of the polyphase filter alone. This is because of their ability to transfer data through the PCIe bus in both directions. However the newest Maxwell generation of gaming GPUs provide the necessary features making them much more appealing, however we note that these cards are unproven in HPC environments. When considering the performance of GPU implementations, the new generation is best as one would expect. However if lower bit precision input data is used the Maxwell generation poses some obstacles. To fully utilise the card capabilities with lower bit precision data one has to use shared memory to avoid superfluous type conversions. We have also provided an analysis of the NVIDIA FFT library, cuFFT, on several generations of hardware. We have demonstrated that using the latest hardware and software might note be the most optimal solution for FFT bound applications on GPUs. 
We have demonstrated acceptable speedups when comparing Xeon Phi to two Intel Xeon CPUs, demonstrating that the Xeon Phi is a useful computations accelerator. We have also noted that development time on different platforms can vary significantly. Meaning that whilst Xeon Phi has a significantly lower performance that our fastest GPU codes, the total cost to a project when considering capital and development expenditure might be comparable. 

	
\section{Acknowledgements}
	The authors would like to acknowledge the following resources used to complete this project:
		\begin{itemize}
		\item{The authors would like to acknowledge the use of the University of Oxford Advanced Research Computing (ARC) facility in carrying out this work. \\ http://dx.doi.org/10.5281/zenodo.22558 }
		\item{The authors would like to acknowledge that the work presented here made use of the Emerald High Performance Computing facility made available by the Centre for Innovation. The Centre is formed by the universities of Oxford, Southampton, Bristol, and University College London in partnership with the STFC Rutherford-Appleton Laboratory.}
		\item{The authors acknowledge the FP7 funded CRISP project, grant agreement number 283745, the Square Kilometre Array (SKA) Telescope Project, the Department of Physics and the Oxford e-Research Centre (OeRC) at the University of Oxford, for use of their system resources and personnel.}
		\item{The authors acknowledge the Oxford University CUDA Centre of Excellence (CCOE) for providing GPU resources.}
		\item The authors acknowledge the internal grants of the Silesian University in Opava FPF SGS/11/2013 and SGS/23/2013.
		\item{The authors would like to thank NVIDIA for providing GPU resources for this project.}
		\item{The authors would like to thank Intel for providing early Xeon Phi hardware for this project.}
		\end{itemize}	

		The authors would also like to thank Mike Giles (UOx), Zden\v ek Stuchl\'ik (Opava), Jayanth Chennamangalam (UOx), and Aris Karastergiou (UOx) for helpful discussions relating to this this work and while preparing this paper.	
	Other software that has been used to complete this project: \textit{MKL -- 11.2.2}, \textit{ICC -- 15.0.2}, \textit{CUDA -- 6.5 (Fermi and Kepler platforms)} and \textit{CUDA -- 7.0 (Maxwell platform)}.


\section{References}
\bibliographystyle{elsarticle-harv} 
\bibliography{polyphase_v2}
	
\end{document}